\documentclass[a4paper,11pt]{article}
\pdfoutput=1 

\usepackage{jcappub} 

\usepackage{multicol}
\usepackage{multirow}
\usepackage[T1]{fontenc} 
\usepackage{graphicx}
\usepackage[export]{adjustbox}
\usepackage{amsmath}
\usepackage{array}
\usepackage{subcaption}
\usepackage{amsfonts}
\usepackage{amssymb}
\usepackage[left=2cm,right=2cm,top=2cm,bottom=2cm]{geometry}
\usepackage{array}
\usepackage{multirow}
\usepackage{multicol}
\usepackage{tablefootnote}
\usepackage{hyperref}
\usepackage[normalem]{ulem} 

\usepackage{xcolor} 
\definecolor{DeepGreen}{RGB}{0,150,150} 

\def\apj{ApJ}

\def\mnras{MNRAS}

\def\aap{A\&A}

\def\nat{Nature}      

\def\apjl{ApJ Letters}

\title {\boldmath Exploring the evolution of red and blue galaxies in different cosmic web environments using IllustrisTNG simulation}

\author[a]{Biswajit Pandey}

\author[a]{and Anindita Nandi}

\affiliation[a]{Department of Physics, Visva-Bharati University, Santiniketan, 731235, India}

\emailAdd{biswap@visva-bharati.ac.in} \emailAdd{anindita.nandi96@gmail.com}

\abstract{We analyze the evolution of red and blue galaxies in
  different cosmic web environments from redshift $z=3$ to $z=0$ using
  the IllustrisTNG simulation. We use Otsu's method to classify the
  red or blue galaxies at each redshift and determine their geometric
  environments from the eigenvalues of the deformation tensor. Our
  analysis shows that initially, blue galaxies are more common in
  clusters followed by filaments, sheets and voids. However, this
  trend reverses at lower redshifts, with red fractions rising earlier
  in denser environments. At $z<1$, most massive galaxies
  ($\log(\frac{M_{*}}{M_{\odot}})>10.5$) are quenched across all
  environments. In contrast, low-mass galaxies
  ($\log(\frac{M_{*}}{M_{\odot}})<10.5$) are more influenced by their
  environment, with clusters hosting the highest red galaxy fractions
  at low redshifts. We observe a slower mass growth for low-mass
  galaxies in clusters at $z<1$. Filaments show relative red fractions
  (RRF) comparable to clusters at low masses, but host nearly $60\%$
  of low-mass blue galaxies, representing a diverse galaxy
  population. It implies that less intense environmental quenching in
  filaments allows galaxies to experience a broader range of
  evolutionary stages. Despite being the densest environment,
  clusters display the highest relative blue fraction (RBF) for
  high-mass galaxies, likely due to interactions or mergers that can
  temporarily rejuvenate star formation in some of them. The $(u-r)$
  colour distribution transitions from unimodal to bimodal by redshift
  $z=2$ across all environments. At $z<1$, clusters exhibit the
  highest median colour, with stellar mass being the primary driver of
  colour evolution in massive galaxies. The suppression of star
  formation rate (SFR) and specific SFR (sSFR) is also most pronounced
  in clusters during this period. Our study suggests that stellar mass
  governs quenching in high-mass galaxies, while a complex interplay
  of mass and environment shapes the evolution of low-mass galaxies.}

\begin{document}
\maketitle
\flushbottom



\section{Introduction}

Galaxy colour bimodality \citep{strateva, blanton03, bell1, balogh,
  baldry04a} has been recognized for more than two
decades. Observations from large galaxy surveys reveal that galaxies
cluster into two distinct groups in colour-magnitude diagrams: a red,
passive population and a blue, star-forming population. The red
galaxies are typically older and have lower star formation
rates. Their red colour indicates that they have largely exhausted
their supply of gas and are composed of older, cooler stars. In
contrast, the blue galaxies are actively forming new stars, which are
hotter. These galaxies are generally younger and have a higher rate of
star formation.

The implications of colour bimodality for galaxy formation and
evolution are significant, as it provides insights into the life cycle
of galaxies. Galaxies follow different evolutionary pathways during
their evolution. The observed dichotomy suggests that galaxies undergo
a transformation from blue to red over time. This transformation may
result from external mechanisms such as interactions, mergers, and
various environmental effects. Alternatively, internal processes,
including gas depletion, stellar evolution, and feedback mechanisms,
may also play crucial roles in this transformation. Understanding how
different internal processes and external influences shape the
observed bimodal colour distribution is essential for gaining a clear
picture of galaxy evolution.

Blue galaxies dominate the cosmic landscape at earlier times due to
the rising gas fractions at higher redshifts. Star formation in these
galaxies is fueled by abundant gas reservoirs. Intense star formation
activity depletes the available gas, leading to a decline in the star
formation rate with time. As star formation gradually ceases with
natural aging, the galaxy eventually becomes red, dominated by older
and cooler stars. In addition to natural aging, various internal
physical mechanisms, such as morphological quenching \citep{martig09},
mass quenching \citep{birnboim03, dekel06, keres05, gabor10}, angular
momentum quenching \citep{peng20}, and bar quenching
\citep{masters10}, can also halt star formation in galaxies. Further,
the expulsion or removal of gas from a galaxy provides another
effective route for quenching star formation. Gas loss caused by
feedback from supernovae, active galactic nuclei, shock-driven winds
\citep{cox04, murray05, springel05a} and ram pressure stripping
\citep{gunn72} can significantly suppress star formation in galaxies.

Besides the internal processes, environment play a crucial role in
transforming blue galaxies into red galaxies. A large body of
literature \citep{hogg04, baldry04b, balogh, blan05, park05, pandey05,
  zehavi05, pandey06, pandey07, bamford09, cooper10, pandey20,
  nandi24} shows that red and blue galaxies tend to inhabit
high-density and low-density regions, respectively. Various
environment-driven mechanisms, such as mergers \citep{hopkins08},
harassment \citep{moore96, moore98}, strangulation \citep{gunn72,
  balogh00}, starvation \citep{larson80, somerville99, kawata08}, and
satellite quenching \citep{geha12}, can inhibit star formation and
alter galaxy structures. Extensive research through simulations
\citep{toomre72,barnes96, mihos96, tissera02, cox06, montuori10,
  lotz11, torrey12, hopkins13, renaud14, renaud15, moreno15, moreno21,
  renaud22, das23} and observations \citep{larson78, barton00,
  lambas03, alonso04, nikolic04, woods06, woods07, barton07,
  ellison08, ellison10, woods10, patton11, barrera15, thorp22, shah22,
  das22} demonstrates that interactions between galaxies, which
produce tidal torques, can trigger starbursts and change both the
colour and morphology of galaxies.

The environment of a galaxy is often characterized by its local
density, but this alone does not describe the neighbourhood of a
galaxy. Galaxies are part of a complex cosmic web composed of
interconnected filaments, clusters, sheets, and voids
\citep{gregory78, joeveer78, einasto80, zeldovich82, einasto84,
  bond96, bharadfil, pandey05, arag10b, libeskind18}, which
collectively organizes and shapes the distribution of matter and
galaxies on a grand scale. Studies with N-body simulations reveal a
dynamic flow of matter within the cosmic web \citep{arag10a, cautun14,
  ramachandra, wang24}, moving from voids to walls, walls to
filaments, and finally into clusters.

Filaments are the largest known coherent structures in the universe
\citep{bharad04, pandey11, sarkar23}. Hydrodynamical simulations
reveal \citep{tuominen21, galarraga21} that more than $40\%-50\%$ of
baryonic matter is found in filaments as the Warm-Hot Intergalactic
Medium (WHIM). This diffuse medium acts as a gas reservoir that can
eventually fall into galaxies, contributing to their growth and
affecting their star formation rates. The presence of WHIM can also
influence gas accretion efficiency in galaxies. Galaxies in different
regions of the cosmic web experience varying levels of gas
accretion. For instance, galaxies near the centers of filaments and
sheets receive a steady supply of cold gas, which fuels star formation
and increases their mass \citep{chen15, pandey20, singh20, das22,
  das23, hoosain24}. Clusters, the densest regions of the universe,
typically form at the intersections of filaments and are characterized
by frequent interactions with neighbouring galaxies and extreme
environmental conditions, leading to relatively rapid transformations
\citep{gunn72, roediger07, ruggiero17}. In contrast, galaxies in
lower-density regions like sheets and voids generally follow quieter
evolutionary paths with subdued star formation \citep{maret22,
  rodriguez24}. Thus, different cosmic web environments play a crucial
role in shaping the diverse evolutionary trajectories of galaxies
across the universe. 

Observations reveal that star formation activity in galaxies peaked
around a redshift of $z\sim 2-3$ \citep{tran10, gupta20}, an era often
dubbed the ``cosmic noon''. The cosmic star formation rate has seen a
dramatic decline from $z=1$ to $z=0$ \citep{madau96}. The number of
massive red galaxies with fixed stellar masses has steadily increased
since $z\sim 1$ \citep{bell04, faber07}. These trends suggest
significant changes in galaxy properties in recent times, which could
be crucial in explaining the observed bimodality in galaxy
distributions. It is essential to understand how the cosmic web
influences the colour bimodality and its evolution.

The IllustrisTNG simulation \citep{nelson18, nelson19, springel18,
  pillepich18, marinacci18, naiman18} models the formation and
evolution of galaxies within the universe. It employs sophisticated
numerical techniques to simulate the complex interactions of dark
matter, gas, and stars across a vast volume of the universe. It
incorporates detailed physics, including gas dynamics, star formation,
supernova feedback, and black hole growth, to provide insights into
galaxy formation and evolution. This simulation is ideal for studying
the roles of different cosmic web environments on galaxy
evolution. Several works \citep{furlong15, trayford16, nelson18,
  wright19, donnari21, walters22, das24} have explored quenching in
galaxies using various hydrodynamical simulations. In the present
work, we will utilize data from the IllustrisTNG simulation to
investigate how different cosmic web environments influence the
transformation of galaxy colour since a redshift of $z\sim 3$.

We will categorize the galaxies into red and blue populations based on
their colour and stellar mass, following a recently introduced
classification scheme \citep{pandey23} based on Otsu's method
\citep{otsu79}.  We will investigate the evolution of the red and blue
fractions in various cosmic web environments. Analyzing these
fractions would help us to assess the efficiency of quenching
processes within each environment. By tracking how the absolute
numbers or percentages of red and blue galaxies change over time, we
can understand how quenching impacts galaxy populations in different
cosmic web environments. Additionally, we will examine the evolution
of the relative red and blue fractions, which reveals how the
proportion of red or blue galaxies in a given environment compares to
those in all other cosmic environments. This comparison will reveal
how the relative abundance of red and blue galaxies in each
environment influences the development of colour bimodality. We will
also explore how these fractions vary with the stellar mass of
galaxies in different cosmic web environments. Studying how the stellar
mass dependence of these fractions changes in different environments
would help us understand the roles of both environment and stellar
mass in shaping the colour bimodality. It is also important to
understand how different galaxy properties evolve in different cosmic
web environments. To explore this, we will examine the evolution of
median colour, median stellar mass, median star formation rate (SFR),
and median specific star formation rate (sSFR) of galaxies with
redshift across various cosmic web environments.

The paper is organized into the following sections. In Section 2, we
provide a detailed overview of the data used in this work. Section 3
explains the methodology employed for the analysis. The findings are
discussed in Section 4, and Section 5 presents the conclusions drawn
from our analysis.

\section{Data}
\label{sec:data}
IllustrisTNG\footnote{\url{https://www.tng-project.org/}}~\cite{marinacci18,
  naiman18, nelson18, springel18, pillepich18, nelson19} is a suite of
cosmological gravo-magnetohydrodynamical simulations built on the
moving-mesh code AREPO~\cite{springel10, weinberger20}. It serves as
an enhanced version of the original Illustris project
\citep{vogelsberger13, vogelsberger14, genel14}. The TNG suite
consists of three cosmological volumes with side lengths of
approximately 50, 100, and 300 Mpc, named TNG50, TNG100, and TNG300,
respectively. Each volume offers multiple resolution levels, such as
TNG100-1, TNG100-2, TNG100-3, and similarly for TNG300, whereas TNG50
has four distinct resolution outputs. All simulations begin at a
redshift of $z = 127$, using cosmological parameters from the Planck
2015 results~\cite{planck15}: $\Omega_{\Lambda} = 0.6911$, $\Omega_m =
0.3089$, $\Omega_b = 0.0486$, $\sigma_8 = 0.8159$, $n_s = 0.9667$, and
$h = 0.6774$. For our analysis, we utilize the TNG300-1 run, which
offers the highest resolution run among the largest available volume
in the TNG series. The specifics of TNG300-1 are detailed in the
~\autoref{tab:tng300_details}, and our study covers redshifts from $z
= 3$ to the present ($z = 0$).

\renewcommand{\arraystretch}{1.5} 
\begin{table}[htbp!]
    \centering
    \begin{tabular}{|c|c|c|c|c|c|c|}
        \hline
        Simulation Name & Volume $(\rm Mpc^3)$ & $N_{GAS}$ & $N_{DM}$ & $m_{baryon}\,(M_{\odot})$ & $m_{DM}\,(M_{\odot})$\\
        \hline
        TNG300-1 & $(302.6)^3$ & $2500^3$ & $2500^3$ & $11\times10^6$ & $59 \times 10^6$ \\ 
        \hline
    \end{tabular}
    \caption{This table describes the specifications of the TNG300-1
      simulation. Different columns in the table represent the
      volume of the box (comoving), initial number of gas cells
      ($N_{GAS}$), number of dark matter particles ($N_{DM}$), the
      target baryon mass ($m_{baryon}$) and dark matter particle mass
      ($m_{DM}$).}
    \label{tab:tng300_details}
\end{table}

At each redshift, we extract various galaxy properties from the
publicly available group catalog, which provides details on both halos
and subhalos. The FoF algorithm is used to identify halos, while the
subhalos (or galaxies) are identified using the SUBFIND
algorithm~\cite{springel01, dolag09}. In this work, we select only
galaxies with a non-zero \textit{SubhaloFlag}, as a value of zero
indicates a subhalo of non-cosmological origin. Additionally, we
impose a stellar mass criterion, choosing galaxies in the range $9 <
\log_{10}(\frac{M_{\star}}{M_{\odot}}) < 12$. Stellar mass is defined
as the total mass of star particles within twice the stellar half-mass
radius $(r_{stars},1/2)$~\cite{pillepich18}. For $(u-r)$ colour, we
use the supplementary catalog provided by \cite{nelson18}, which
provides apparent magnitudes for different snapshots and
dust-corrected absolute magnitudes only at $z=0$. At $z=0$, we use the
dust-corrected absolute magnitudes to compute colours. For all other
redshifts ($z > 0$), we compute colours using the apparent magnitudes,
as dust-corrected values are not available. Since colour is determined
by the difference between magnitudes in two bands, and the distance
modulus is the same for both bands of a given galaxy, the colour
remains a relative quantity. Although the absence of dust correction
at $z>0$ may introduce systematic offsets, our classification method
is designed to be robust to such effects. Specifically, we apply
Otsu’s thresholding method \citep{otsu79, pandey23} to the $(u-r)$
colour distribution at each redshift and in different stellar mass
bins. This data-driven method identifies the optimal colour threshold
based on the bimodal distribution of galaxy colours, allowing for a
consistent and redshift-dependent separation of red and blue
populations without relying on fixed colour cuts or rest-frame
corrections. The bimodal distribution is more prominent in $(u-r)$
than other colours, such as $(g-r)$. The u-band flux is strongly
affected by young, hot stars, while r-band flux traces older stars.
This makes $(u-r)$ colour a better indicator of recent star formation
activity, whereas $(g-r)$ is less sensitive to changes in young
stellar populations.  We obtain the star formation rates from the
\textit{SubhaloSFRinRad} field in the group catalog. In IllustrisTNG,
star formation is modeled following \cite{springel03}. For our
analysis, we use the instantaneous star formation rate, which is the
sum of the star formation rates of all star-forming gas cells within
twice the stellar half-mass radius of the galaxy.

\begin{figure}[htbp!]
\centering
\includegraphics[width = 1.0\textwidth]{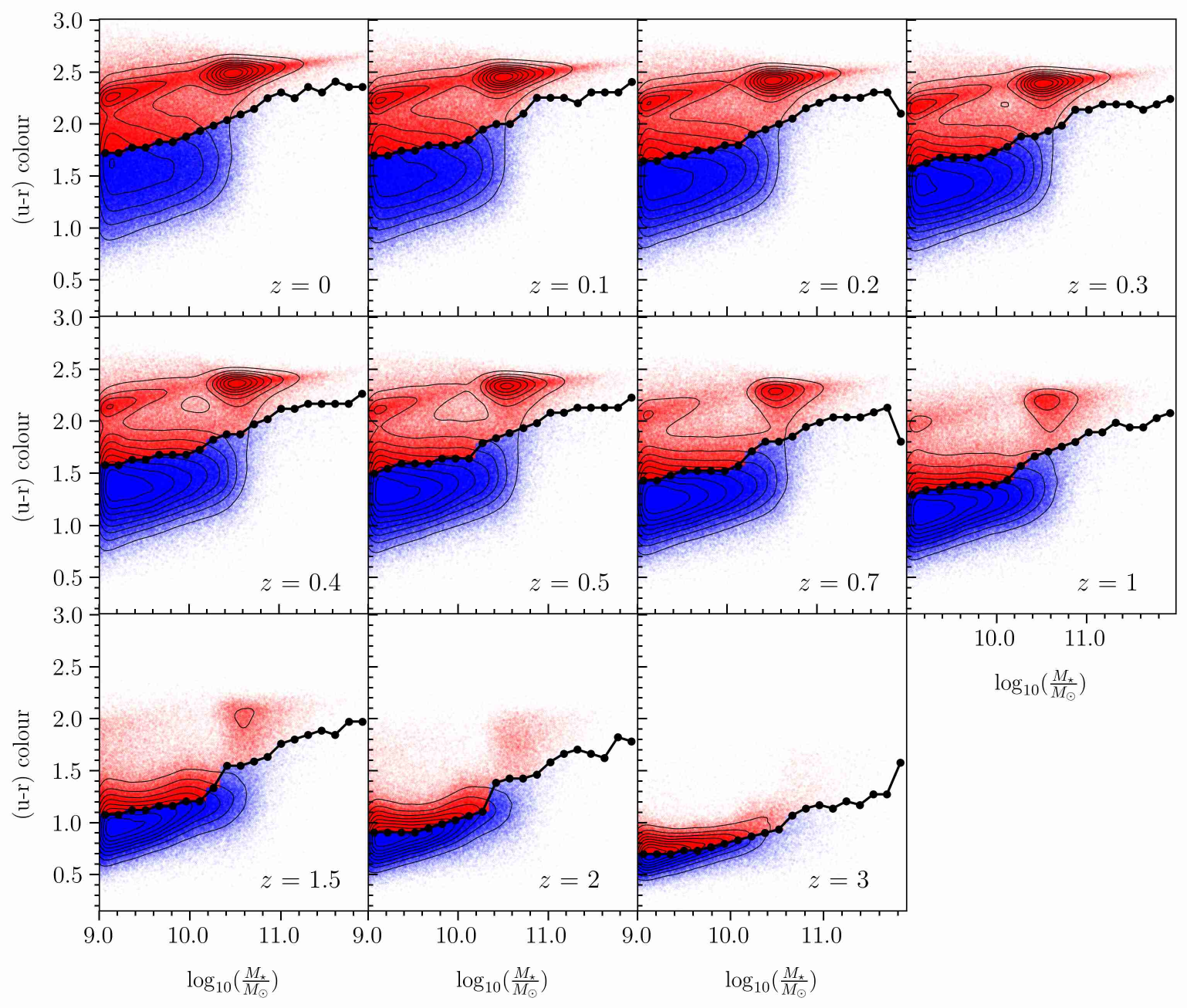}
\caption{This figure shows the distribution of galaxies in the $(u-r)$
  colour-stellar mass plane across each redshift, with the dividing
  line (determined by Otsu's method) distinguishing the red and blue
  populations. In each subplot, we also show $10$ uniformly spaced
  contours lying between the minimum and maximum density of galaxies
  in the respective colour-stellar mass plane.}
 \label{fig:otsu}
\end{figure}

\section{Method}
The primary objective of this study is to classify galaxies in the
IllustrisTNG simulation into red and blue populations according to
their stellar mass and redshift, and to investigate how cosmic web
environments and stellar mass influence their evolution.

The $(u-r$) colour distribution is bimodal at low redshift, with peaks
corresponding to the blue cloud and the red sequence. Traditionally a
fixed $(u-r)$ colour cut is used in many earlier studies to separate
the red and blue galaxies. But a fixed colour threshold assumes the
same division between red and blue galaxies regardless of redshift or
stellar mass, which does not reflect galaxy evolution. The precise
location of the peaks of the two populations evolves with redshift and
stellar mass. Thus, a static threshold could misclassify galaxies,
particularly at higher redshifts where the colour distribution is less
bimodal. Otsu’s method \citep{otsu79, pandey23} is better than a fixed
colour cut because it provides adaptive, optimal separation of red and
blue galaxies while accounting for variations across different stellar
mass bins and redshifts.

\subsection{Identifying red and blue Galaxies with Otsu's method}\label{sec:method1}

Otsu’s thresholding technique \citep{otsu79}, originally designed for
image segmentation, separates data into two groups by minimizing the
intra-class variance (variance within each group) and maximizing the
inter-class variance (variance between groups). This method iterates
through all possible thresholds, selecting the one that provides the
optimal separation between the two classes. This technique is
particularly useful when the data distribution is bimodal, as is the
case with the $(u-r)$ colour distribution of galaxies, where we
observe two peaks corresponding to the blue cloud and the red
sequence. Recently, \cite{pandey23} proposed a method for classifying
galaxies into two groups: the ``blue cloud'' (younger, star-forming
galaxies) and the ``red sequence'' (older, passive galaxies) based on
the Otsu's technique.

The primary steps in this method are as follows.
\begin{enumerate}
\item Histogram Calculation: We first calculate the histogram of the
  $(u-r)$ colour for all galaxies in our sample, using a specific
  number of bins $M$. The histogram is then normalized by the total
  number of galaxies $N=\sum_{i=1}^{M} n_i$ where $n_i$ corresponds to
  the number of galaxies in the $i^{th}$ colour bin. This ensures that
  the total probability across all bins sums to one
  i.e. $\sum_{i=1}^{M} p_{i}=1$ where $p_{i}=\frac{n_{i}}{N}$.

  \item Class Probabilities: A threshold ($k$-th bin) is chosen,
    dividing galaxies into blue cloud (bins $1$ to $k$) and red
    sequence (bins $k+1$ to $M$). We iterate through all possible
    thresholds (i.e. $k$ assumes all values between $1$ to $M-1$). The
    probability of a galaxy belonging to each class is computed. The
    class probabilities of the blue cloud (BC) and the red sequence
    (RS) are respectively given by,

  \begin{eqnarray}
  P_{BC}=\sum_{i=1}^{k} p_{i}=w(k)
\label{eq:weights1}
\end{eqnarray}
and
\begin{eqnarray}
  P_{RS}=\sum_{i=k+1}^{M} p_{i}=1-w(k).
\label{eq:weights2}
\end{eqnarray}

\item Class means and Class variances: We calculate the mean colour
  and variance for both groups for each threshold value.

The means of the blue cloud and the red sequence are calculated as,\\

\begin{eqnarray}
  \mu_{BC}=\frac{\sum_{i=1}^{k} x_{i} p_{i}}{P_{BC}}=\frac{\mu_{k}}{w(k)}
\label{eq:mu1}
\end{eqnarray}
and
\begin{eqnarray}
  \mu_{RS}=\frac{\sum_{i=k+1}^{M} x_{i} p_{i}}{P_{RS}}=\frac{\mu_{T}-\mu_{k}}{1-w(k)}
\label{eq:mu2}
\end{eqnarray}
Here, $x_{i}$ represents the $(u-r)$ colour corresponding to the
$i^{th}$ bin. The mean up to the $k^{th}$ bin is given by
$\mu_{k}=\sum_{i=1}^{k} x_{i} p_{i}$, while $\mu_{T}=\sum_{i=1}^{M}
x_{i} p_{i}$ denotes the mean of the entire distribution. It is
important to note that $P_{BC}+P_{RS}=1$ and that
$\mu_{T}=P_{BC}\,\mu_{BC}+P_{RS}\,\mu_{RS}$ for every threshold.

Similarly, the class variances are estimated as,\\

\begin{eqnarray}
  \sigma_{BC}^{2}=\frac{\sum_{i=1}^{k} (x_{i}-\mu_{BC})^2 p_{i}}{P_{BC}}
\label{eq:sigma1}
\end{eqnarray}
and
\begin{eqnarray}
  \sigma_{RS}^{2}=\frac{\sum_{i=k+1}^{M} (x_{i}-\mu_{RS})^2 p_{i}}{P_{RS}}
\label{eq:sigma2}
\end{eqnarray}

\item Intra-class variance and Inter-class variance: The total
  variance is divided into intra-class (within-class) variance and
  inter-class (between-class) variance. The intra-class variance
  ($\sigma_{wc}^2$) and the inter-class variance ($\sigma_{bc}^2$) are
  calculated as,
\begin{eqnarray}
   \sigma_{wc}^2=P_{BC}\, \sigma_{BC}^{2}+P_{RS}\, \sigma_{RS}^{2}
\label{eq:intra}
\end{eqnarray}
and
\begin{eqnarray}
   \sigma_{bc}^2=P_{BC}\,P_{RS}\,(\mu_{BC}-\mu_{RS})^2
\label{eq:inter}
\end{eqnarray}

The total variance $\sigma_{T}^2$ is the sum of the two variances,
\begin{eqnarray}
\sigma_{T}^2=\sigma_{wc}^2+\sigma_{bc}^2
\label{eq:total}
\end{eqnarray}
It is important to note that both $\sigma_{wc}^2$ and $\sigma_{bc}^2$
are dependent on the selected threshold, while $\sigma_{T}^2$ remains
independent of it.

\item Threshold Selection: The optimal threshold is the one that
  minimizes the intra-class variance $\sigma_{wc}^2$, which is
  equivalent to maximizing the inter-class variance
  $\sigma_{bc}^2$. We iterate through all possible thresholds,
  selecting the one that minimizes the intra-class variance while
  maximizing the inter-class variance. This threshold is not affected
  by binning choices, making it a reliable approach for classifying
  red and blue galaxies \citep{pandey23}.
  
\end{enumerate}

\begin{table}[htbp!]
\centering
\begin{tabular}{|c|c|c|c|c|c|c|c|c|c|}
\hline
\multirow{3}{*}{Redshift} &  \multicolumn{9}{|c|}{Number of galaxies} \\
\cline{2-10}

 &  \multicolumn{3}{|c|}{All} &  \multicolumn{3}{|c|} {$9 < \log_{10}\left(\frac{M_{\star}}{M_{\odot}}\right) \leq 10.5$} &  \multicolumn{3}{|c|}{ $10.5 < \log_{10}\left(\frac{M_{\star}}{M_{\odot}}\right) < 12$} \\
\cline{2-10}
 & Total & Red & Blue & Total & Red & Blue &  Total & Red & Blue \\
\hline
$0$ & $253140$ & $130751$ & $122389$ & $225075$ & $105417$ & $119658$ & $28065$ & $25334$ & $2731$ \\
\hline
$0.1$ & $253212$ & $120682$ & $132530$ & $225135$ & $96243$  & $128892$ & $28077$ & $24439$ & $3638$ \\
\hline
$0.2$ & $252546$ & $112362$ & $140184$ & $224612$ & $89094$ & $135518$ & $27934$ & $23268$ & $4666$ \\
\hline
$0.3$ & $251329$ & $105559$ & $145770$ & $223692$ & $83375$ & $140317$ & $27637$ & $22184$ & $5453$ \\
\hline
$0.4$ & $249659$ & $96073$ & $153586$ & $222588$ & $75211$ & $147377$ & $27071$ & $20862$ & $6209$ \\
\hline
$0.5$ & $247760$ & $90513$ & $157247$ & $221341$ & $71415$ & $149926$ & $26419$ & $19098$ & $7321$ \\
\hline
$0.7$ & $242859$ & $81067$ & $161792$ & $218375$ & $64601$ & $153774$ & $24484$ & $16466$ & $8018$ \\
\hline
$1$ & $231933$ & $73719$ & $158214$ & $210682$ & $60488$ & $150194$ & $21251$ & $13231$ & $8020$ \\
\hline
$1.5$ & $199307$ & $69796$ & $129511$ & $184161$ & $61093$ & $123068$ & $15146$ & $8703$ & $6443$ \\
\hline
 $2$ & $157394$ & $66878$ & $90516$ & $147377$ & $61740$ & $85637$ & $10017$ & $5138$ & $4879$ \\
 \hline
 $3$ & $84438$ & $46230$ & $38208$ & $80992$ & $44559$ & $36433$ & $3446$ & $1671$ & $1775$ \\
 \hline
\end{tabular}
\caption{This table shows the total number of galaxies with nonzero
  \textit{SubhaloFlag} and stellar masses within $9 <
  \log_{10}(\frac{M_{\star}}{M_{\odot}}) < 12$ in IllustrisTNG
  simulation for each snapshot from redshift $0$ to $3$. It also lists
  the number of red and blue galaxies identified with the Otsu's
  method at each redshift in different mass ranges.}
\label{tab:otsunumber}
\end{table}

There are distinct relationships between colour and stellar mass or
absolute magnitude. Therefore, a single colour threshold cannot be
justified for galaxies of varying masses and luminosities. We divide
the entire sample into several independent stellar mass bins and apply
this technique to each of these bins separately, allowing the
classification to adapt to how colour varies with mass. This provides
us a dividing line between the two populations in the colour-stellar
mass plane at each redshift (\autoref{fig:otsu}). Otsu's method
determines the optimal division between red and blue population at
each redshift by minimizing the intra-class variance and maximizing
the inter-class variance, ensuring the separation is statistically
robust.

We tabulate the number of red and blue galaxies in two different mass
ranges between redshift $3$ to $0$ in \autoref{tab:otsunumber}.

It may be noted that at higher redshifts, particularly at $z = 3$, the
$(u-r)$ colour distribution becomes less clearly bimodal (bottom right
panel of \autoref{fig:colpdf}). One-dimensional histograms at this
epoch show a largely unimodal shape, making the identification of
distinct red and blue populations less straightforward. However, when
viewed in the stellar mass-colour plane, a subtle population of
massive galaxies with relatively higher $(u-r)$ colours emerges,
suggesting that a weak form of bimodality may already be developing
(bottom right panel of \autoref{fig:otsu}). To maintain methodological
consistency across redshifts, we continue to apply Otsu’s thresholding
method at $z = 3$ as a data-driven way to divide the galaxy population
into two relative colour classes. At this redshift, the resulting
groups are better interpreted as ``less blue'' and ``more blue''
galaxies, rather than physically distinct red and blue
populations. This functional classification still provides meaningful
insights into environmental trends at early times. This approach
allows for a uniform comparative analysis across cosmic time, while
acknowledging the evolving nature of galaxy colour distributions.

\begin{figure}[htbp!]
\centering
\includegraphics[width = 0.75\textwidth]{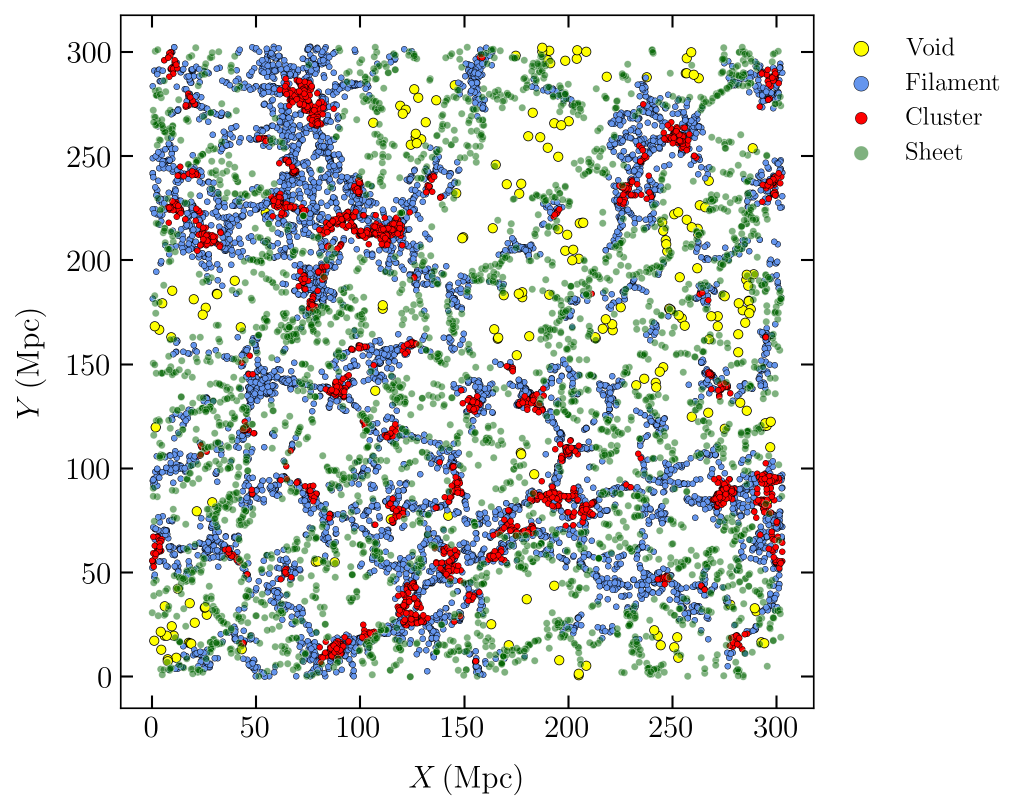}
\caption{This figure displays the distribution of galaxies in various
  cosmic web environments at $z=0$, from a 20 Mpc thick slice of the
  IllustrisTNG simulation. The cosmic web classification is based on
  the eigenvalues of the deformation tensor, estimated from the
  density field smoothed with a Gaussian filter of width 4 Mpc.}
 \label{fig:hess}
\end{figure}

\subsection{Identifying different morphological environments of the cosmic web}\label{sec:method2}
We classify galaxies into different morphological environments within
the cosmic web using a Hessian-based method \citep{hahn2,
  fromero}. This method relies on the eigenvalues and eigenvectors of
the deformation tensor to identify voids, sheets, filaments, and
clusters.

The deformation tensor $T_{ij}$ is derived from the Hessian matrix of
the gravitational potential field $\Phi$, defined as

\begin{equation}
T_{ij} = \frac{\partial^2 \Phi}{\partial x_i \partial x_j}
\end{equation}
where $x_i$ and $x_j$ are the spatial coordinates. The gravitational
potential $\Phi$ is computed by solving the Poisson equation

\begin{equation}
\nabla^2 \Phi \equiv \delta
\end{equation}

Here, $\delta = \frac{\rho - \bar{\rho}}{\bar{\rho}}$ is the density
contrast, with $\rho$ representing the local density and $\bar{\rho}$
the average density. To calculate the potential, we apply the
Cloud-In-Cell (CIC) scheme to construct a discrete density contrast
field on a grid. This field is smoothed using an isotropic Gaussian
filter with a width of 4 Mpc. The mean intergalactic separation of our
samples varies between $4.79$ Mpc (at $z=0$) to $6.91$ Mpc (at
$z=3$). Further, we have calculated the potential using a discretized
density field with a grid spacing of $\sim 1.18$ Mpc. The smoothing
scale should be larger than the grid spacing to ensure meaningful
results. We are primarily concerned about resolving large-scale
structures of the cosmic web.  So, we have chosen a filter width of 4
Mpc for our present analysis. We also study the effects of varying the
filter width in \autoref{subsec:diffsmooth}.

Next, we compute the Fourier transform of the gravitational potential:

\begin{equation}
\hat{\Phi} = \hat{G} \hat{\rho}
\end{equation}

where $\hat{G}$ is the Fourier transform of the Green’s function of
the Laplacian operator, and $\hat{\rho}$ is the density in Fourier
space. By transforming the potential back into real space, we
calculate the tidal tensor using numerical differentiation.

Based on the signs of the three eigenvalues $\lambda_1$, $\lambda_2$,
and $\lambda_3$ (where $\lambda_1 > \lambda_2 > \lambda_3$), galaxies
are classified into different cosmic web environments as follows:

\begin{enumerate}
    \item Void: $\lambda_1, \lambda_2, \lambda_3 < 0$
    \item Sheet: $\lambda_1 > 0$, $\lambda_2, \lambda_3 < 0$
    \item Filament: $\lambda_1, \lambda_2 > 0$, $\lambda_3 < 0$
    \item Cluster: $\lambda_1, \lambda_2, \lambda_3 > 0$
\end{enumerate}

Each type represents a distinct geometric environment in the cosmic
web, with voids being underdense and clusters representing the densest
regions.

It may be worthwhile to mention that the density field used in this
study is based on the spatial distribution of galaxies, rather than
the underlying dark matter field. While the large-scale structure is
fundamentally shaped by the dark matter distribution, and galaxies are
biased tracers of that field, our choice reflects the practical
constraints of observational studies. Since dark matter is not
directly observable, galaxy-based density fields provide a more
realistic and consistent basis for comparison with observational
data. Our focus on galaxy properties motivates the use of a density
measure derived from the same population, facilitating a more direct
link between environment and observable galaxy characteristics. We
utilize a subset of subhalos in each redshift applying the criteria
discussed in \autoref{sec:data}. The total number of subhalos in each
redshift are presented in the first sub-column (``Total'') of the
first column (``All'') of \autoref{tab:otsunumber}.

\begin{table}[htbp]
\centering
\begin{tabular}{|c|c|c|c|c|c|}
\hline
{\rule{0pt}{4ex} Redshift} & \multicolumn{5}{|c|}{Number of Galaxies} \\ \cline{2-6}
 & Total & Void & Sheet & Filament & Cluster \\
\hline
0 & 253140 & 1742 & 40510 & 128945 & 81943 \\ 
\hline
0.1 & 253212 & 1746 & 40213 & 129829 & 81424 \\ 
\hline
0.2 & 252546 & 1728 & 39965 & 130091 & 80762 \\ 
\hline
0.3 & 251329 & 1729 & 39453 & 129785 & 80362 \\ 
\hline
0.4 & 249659 & 1749 & 38779 & 129826 & 79305 \\ 
\hline
0.5 & 247760 & 1725 & 38092 & 128924 & 79019 \\ 
\hline
0.7 & 242859 & 1668 & 36612 & 127139 & 77440 \\ 
\hline
1 & 231933 & 1580 & 34128 & 121051 & 75174 \\ 
\hline
1.5 & 199307 & 1187 & 27280 & 103015 & 67825 \\ 
\hline
2 & 157394 & 850 & 19497 & 80358 & 56689 \\ 
\hline
3 & 84438 & 224 & 7860 & 41229 & 35125 \\ 
\hline
\end{tabular}
\caption{This table represents the number of galaxies identified in
  the different regions of the cosmic web at different redshifts.}
\label{tab:galaxies_in_web}
\end{table}

\begin{figure}
    \centering
    \includegraphics[width = 0.9\textwidth]{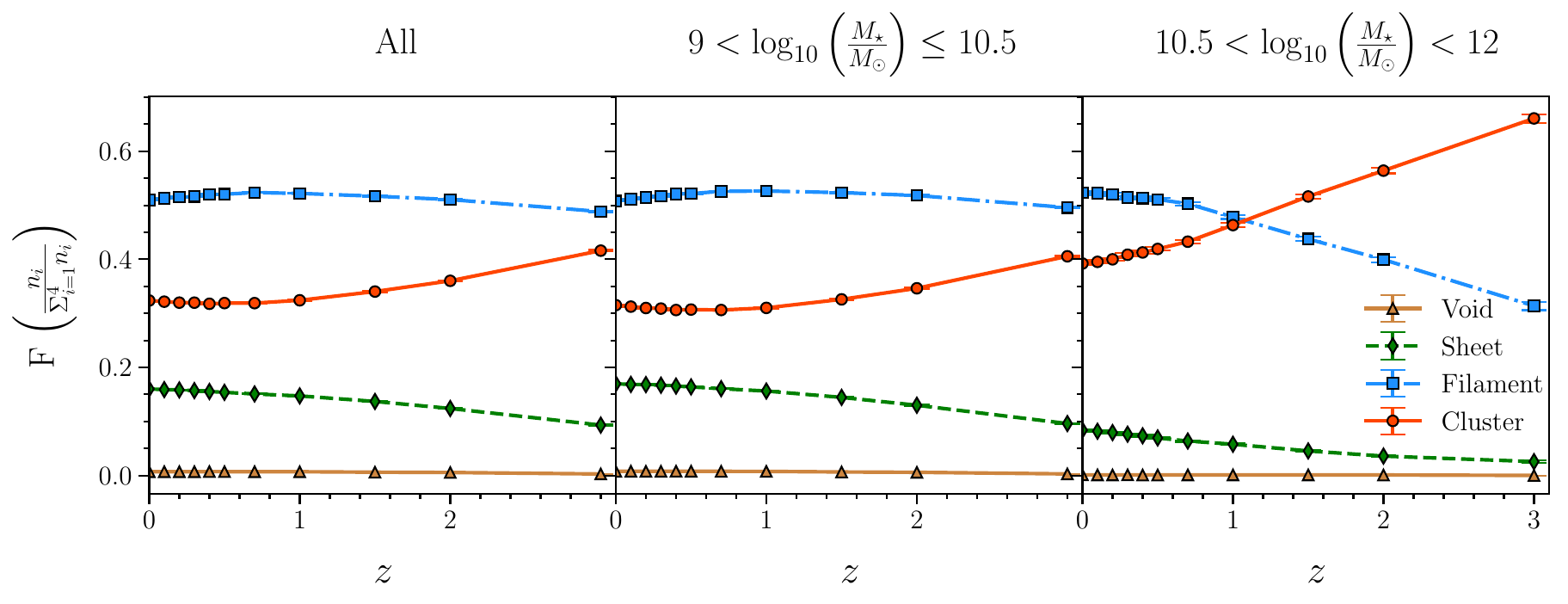}
    \caption{The panels in this figure show the evolution of the
      relative fraction of galaxies in different cosmic web
      environments with redshift for different mass ranges. The
      1$\sigma$ Binomial errorbars are shown at each data point.}
    \label{fig:frac}
\end{figure}

\begin{figure}[htbp!]
\centering
\includegraphics[width = 1.0\textwidth]{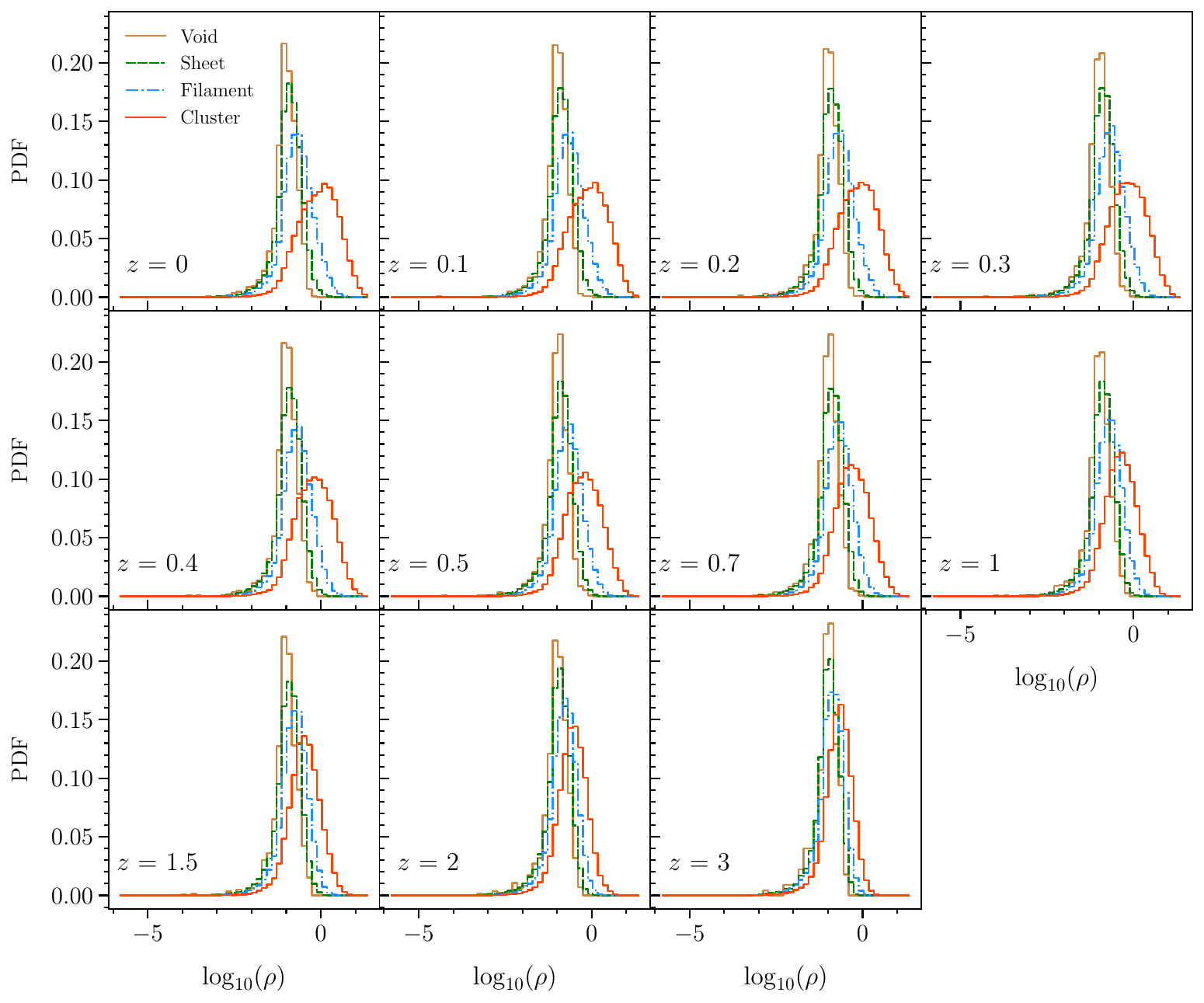}
\caption{This figure shows the distribution of galaxy local densities
  across different cosmic web environments at each redshift. Here,
  local density refers to the density estimated using the CIC
  interpolation scheme applied to the galaxy distribution.}
\label{fig:density}
\end{figure}

We list the number of galaxies present in each type of cosmic web
environments at different redshifts in
\autoref{tab:galaxies_in_web}. We also show the evolution of the
relative fraction of galaxies in each type of cosmic web environments
for the entire sample and the two distinct mass bins in
\autoref{fig:frac}.  The results are obtained for a smoothing length
of 4 Mpc. The left panel of \autoref{fig:frac} shows that filaments
generally have the highest galaxy abundance, followed by clusters,
sheets, and voids across the entire redshift range. The proportion of
galaxies in filaments and sheets increases with decreasing redshift,
while the proportion in clusters decreases. These trends are also
evident in the lower mass bin, as depicted in the middle panel of
\autoref{fig:frac}. In the high mass bin, the proportion of galaxies
in filaments and clusters evolves differently, as shown in the right
panel of \autoref{fig:frac}. There is a crossover around $z\sim 1$,
after which filaments have a higher proportion of galaxies compared to
clusters. Prior to $z=1$, clusters dominate in terms of galaxy
proportion.

An earlier analysis \citep{cautun14} using the Millennium simulation
\citep{springel05b} find that in the present Universe, the mass
fraction in filaments and sheets are $\sim50\%$ and $\sim 24\%$
respectively. Another study \citep{veena19} with EAGLE simulation
\citep{schaye15} find that $\sim 62-66 \%$ galaxies with stellar mass
larger than 5 times $10^8\, M_{\odot}/h$ are in filaments at
$z=0$. Despite differences in the method, sample and simulation, it is
interesting to note that our analysis indicates $\sim 52\%$ galaxies
are in filaments and $\sim 18\%$ galaxies are in sheets at $z=0$.

We also show the probability distribution function (PDF) of the local
density across various cosmic web environments at different redshifts
in different panels of \autoref{fig:density}. The peaks of the
distributions for different cosmic web environments occur at different
densities. The peak for nodes appears at the highest density, followed
in order by filaments, sheets, and voids. At each redshift, the low
density regions are dominated by voids followed by sheets and
filaments. The peak for the nodes exhibit a gradual shift towards
higher densities with decreasing redshift.

\section{Results}
\label{sec:results}
In this section we will study the the red fraction (RF) and blue
fraction (BF) in different cosmic web environments as a function of
redshift and stellar mass. The red fraction (RF) and blue fraction
(BF) in any cosmic web environment are respectively defined as
RF$=\frac{n_R}{n_R+n_B}$ and BF$=\frac{n_B}{n_R+n_B}$.  Here, $n_R$
and $n_B$ are the number of red and blue galaxies in the same
environment. The RF and BF thus provide the proportion of two
populations in any particular environment. We will also analyze the
relative red fraction (RRF) and relative blue fraction (RBF) in
different parts of the cosmic web as a function of redshift and
stellar mass. We can use the relative red fraction (RRF) and the
relative blue fraction (RBF) to study the abundance of red and blue
galaxies in a given environment relative to the total cosmic web. The
RRF and RBF are respectively defined as
RRF$=\frac{(n_R)_i}{\sum_{i=1}^4 (n_R)_i}$ and
RBF$=\frac{(n_B)_i}{\sum_{i=1}^4 (n_B)_i}$, where $(n_R)_i$ and
$(n_B)_i$ are the numbers of red and blue galaxies in $i^{th}$-type
environment. Here $i$ runs from $1$ to $4$ covering 4-types of cosmic
web environment.

It may be noted that the red fraction (RB) and blue fraction (BF) are
complementary to each other, meaning that their sum always equals
1. However, this relationship does not extend to the relative red
fraction (RRF) and relative blue fraction (RBF). In other words,
knowing the RRF does not allow us to directly infer the RBF.

\subsection{Evolution of red and blue fractions in different cosmic web environments}
\label{subsec:rbfrac}

\begin{figure}
    \centering
    \includegraphics[width = 0.9\textwidth]{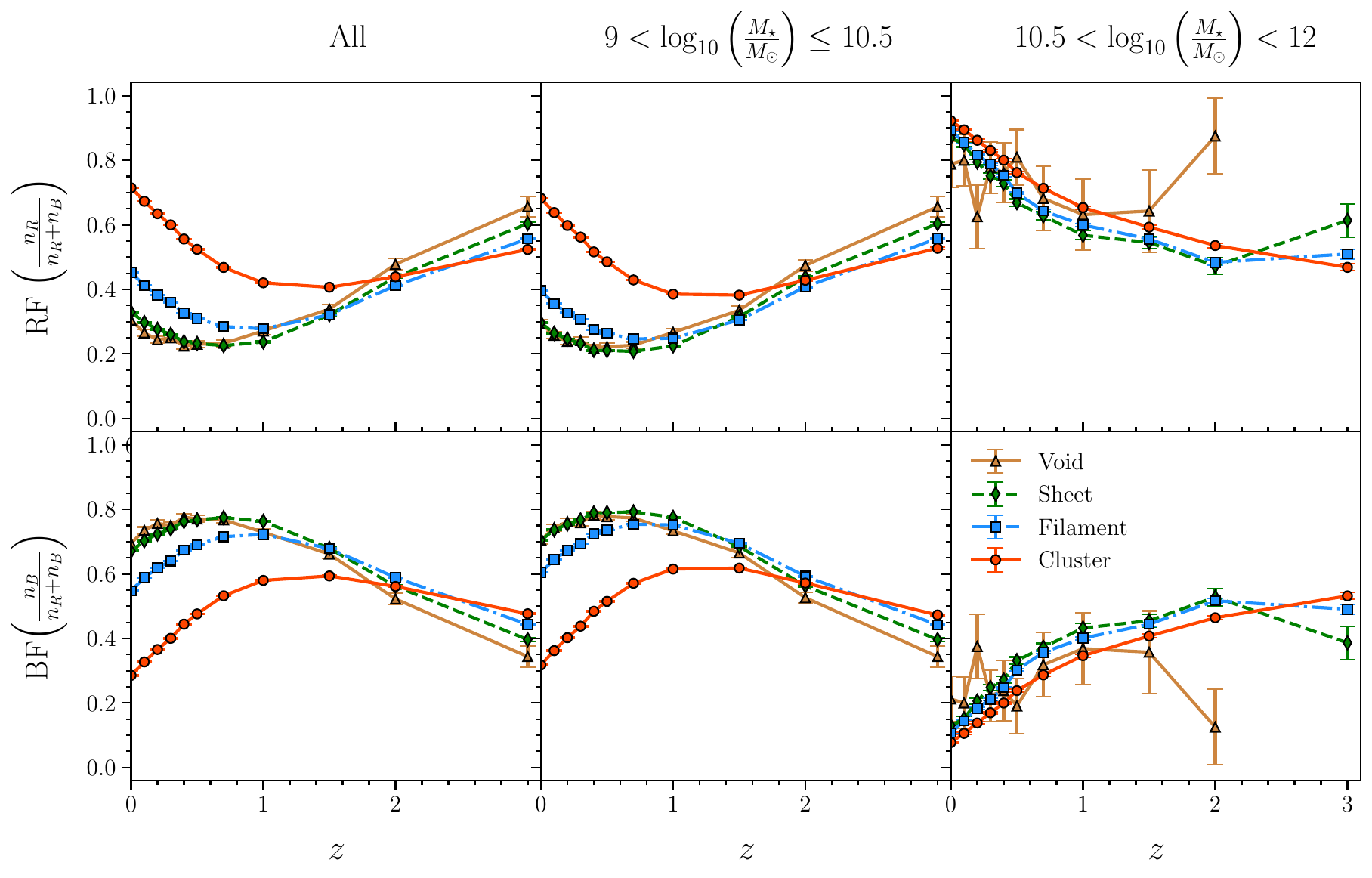}
    \caption{Different panels in this figure display the red and blue
      fractions across various cosmic web environments, plotted as a
      function of redshift for different mass ranges. The 1$\sigma$
      Binomial errorbars are shown at each data point.}
    \label{fig:rbfrac}
\end{figure}

In this subsection, we examine the evolution of the fraction of red
and blue galaxies in various environments within the cosmic web. The
top left and bottom left panels of \autoref{fig:rbfrac} illustrate the
red and blue fractions (RF and BF) across different cosmic web
environments. Although the blue fraction can be obtained by simply
subtracting the red fraction from one, they are separately shown here
for the sake of completeness.

From the bottom left panel of \autoref{fig:rbfrac}, we observe that at
redshift $z=3$, the blue fraction is highest in clusters, followed by
filaments, sheets, and voids. Specifically, the blue fraction is $\sim
48\%$ in clusters and around $35\%$ in voids. Although the differences
in blue fraction across environments at $z=3$ are not substantial, the
trend suggests that overdense regions were more conducive to star
formation in earlier times. This may arise due to the presence of
larger gas reservoirs in these regions.

The first column of \autoref{fig:rbfrac} shows that as redshift
decreases, the blue fraction increases in all cosmic web
environments. However, this trend reverses at lower redshifts, with
the blue fraction declining and the red fraction rising. The blue
fraction decreases and the red fraction increases at specific
redshifts, with inflection points occurring at $z=1.5$ for clusters,
$z=0.7$ for filaments and sheets, and $z=0.5$ for voids. At
$z\leq1.5$, red fractions dominate in clusters, followed by filaments,
sheets, and voids. By $z=0$, the red fraction becomes $75\%$ in
clusters, $45\%$ in filaments, $34\%$ in sheets, and $30\%$ in voids.

These results indicate an environment-dependent trend in galaxy colour
transformation. At $z = 3$, clusters exhibit the highest blue galaxy
fraction, followed by filaments, sheets, and voids. This trend
suggests that early clusters may have hosted more star-forming
galaxies, potentially due to higher gas availability. The blue
fraction increases across all environments until redshift $z = 1.5$,
which likely corresponds to the peak of cosmic star formation
activity. After $z = 1.5$, the blue fraction begins to decline, with
red galaxies becoming dominant in clusters. This shift may reflect the
onset of processes that suppress star formation, such as gas
depletion, AGN feedback, or galaxy interactions, which are expected to
be more prevalent in dense environments. The timing of this transition
varies across environments, with filaments and sheets showing later
changes compared to clusters. This pattern suggests that the influence
of environment on galaxy evolution may depend on local density,
although we caution that our analysis does not establish
causality. The blue fraction remains highest in voids until $z < 0.5$,
after which an increase in red galaxy fraction is observed. These
trends are consistent with the idea that quenching processes become
effective earlier in denser environments. The inflection points seen
in \autoref{fig:rbfrac} thus reflect a sequence in galaxy evolution
that aligns with increasing environmental density, though we emphasize
that these are observed trends and not direct evidence of specific
physical mechanisms.

\begin{figure}
    \centering
    \includegraphics[width = 0.9\textwidth]{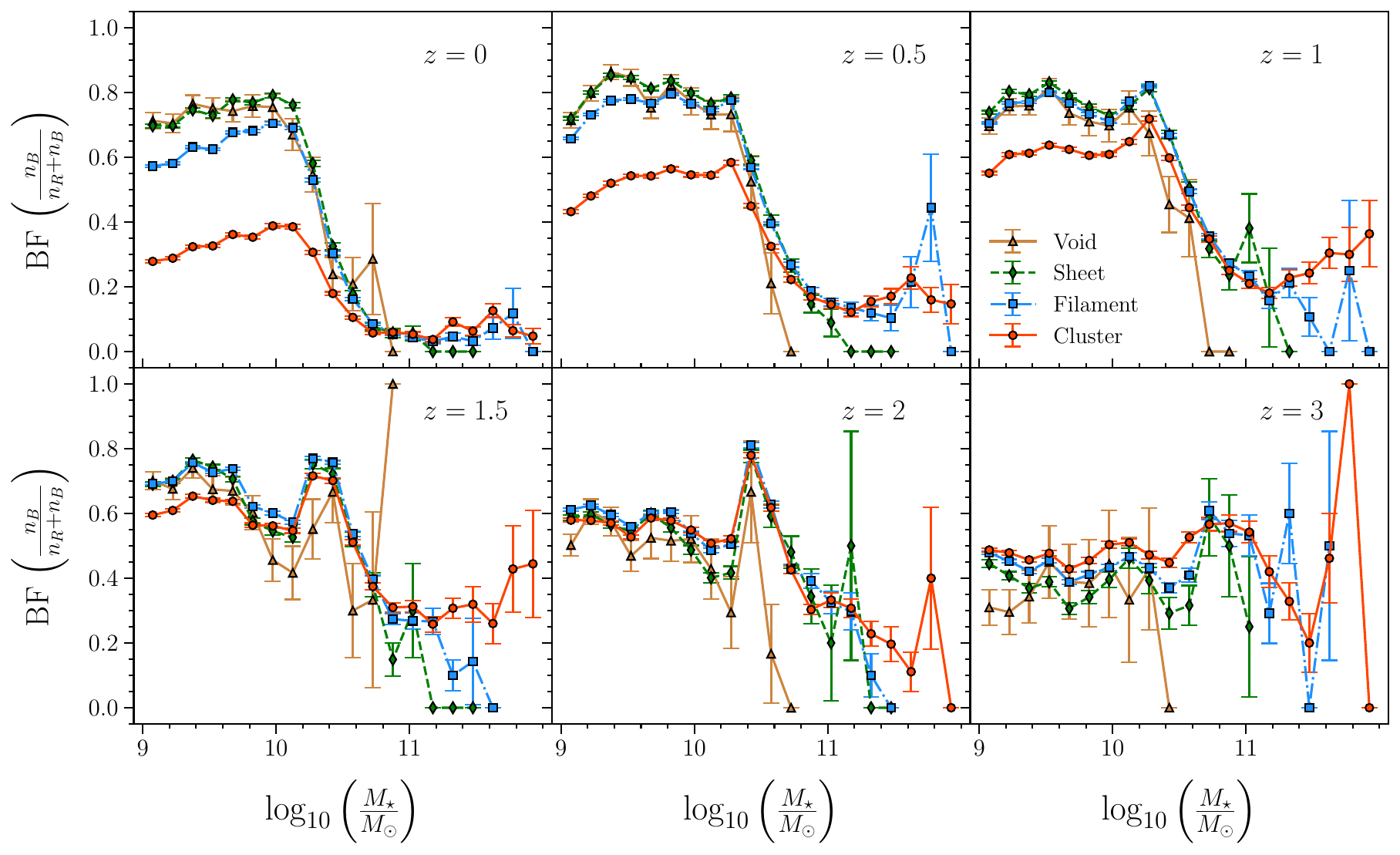}
    \caption{The panels of this figure show the blue
        fraction as a function of stellar mass in different cosmic web
        environments at different redshifts. The 1$\sigma$ Binomial
        errorbars are shown at each data point.}
    \label{fig:bfracmass}
\end{figure}

\begin{figure}
    \centering
    \includegraphics[width = 0.9\textwidth]{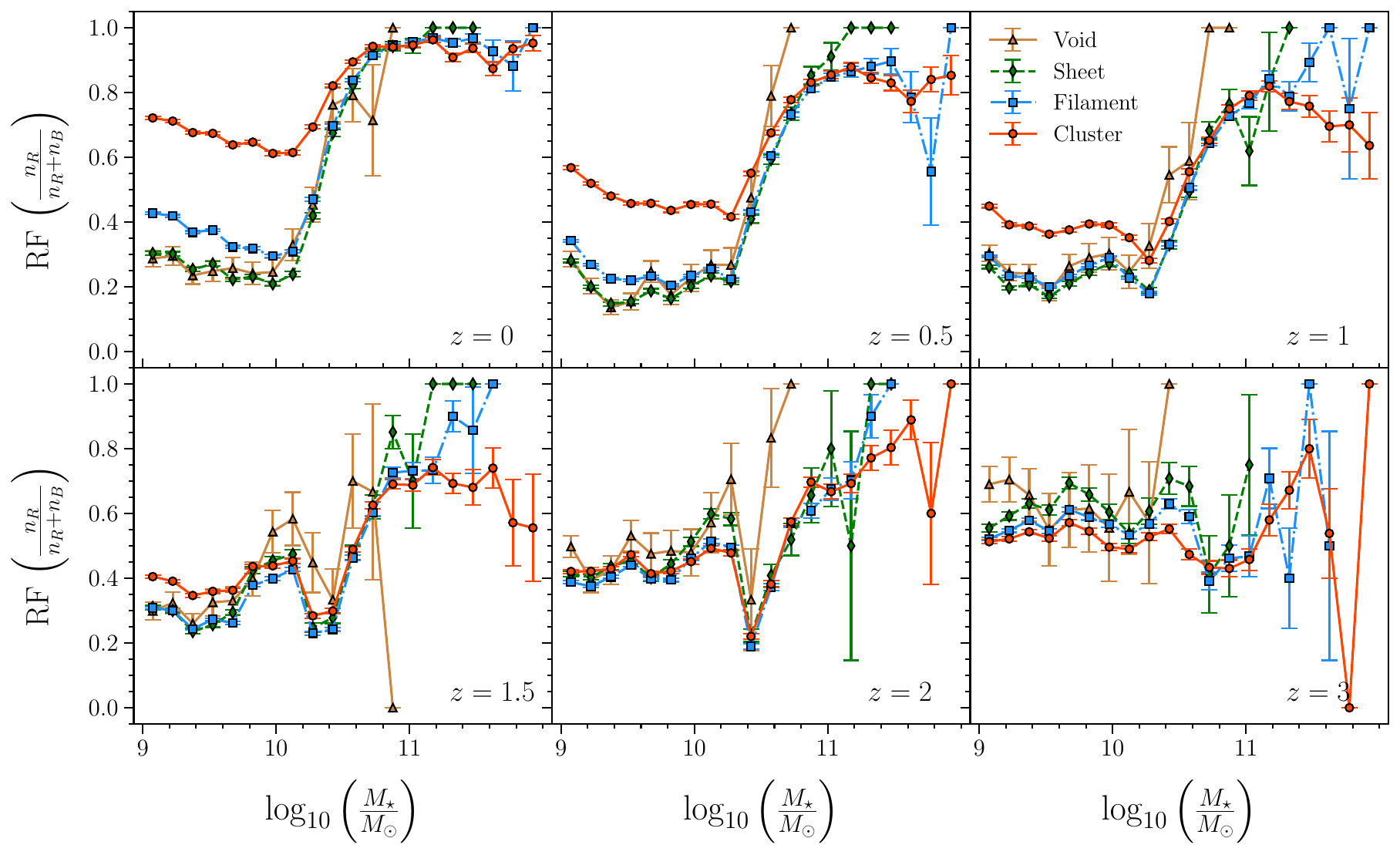}
    \caption{Same as \autoref{fig:bfracmass} but for red fraction.}
    \label{fig:rfracmass}
\end{figure}

We now divide our galaxy sample at each redshift into two distinct
mass bins: $9<\log(\frac{M_{*}}{M_{\odot}})\leq 10.5$ and
$10.5<\log(\frac{M_{*}}{M_{\odot}})<12$ to examine the evolution of
red and blue fractions in lower and higher mass galaxies
separately. Results for these two mass bins are presented in the
second and third columns of \autoref{fig:rbfrac}. The second column of
\autoref{fig:rbfrac} displays the evolution of red and blue fractions
for the lower mass bin, which includes approximately $\sim 90\%$
galaxies of the entire sample (\autoref{tab:otsunumber}). For this
reason, the results for the lower-mass bin closely resemble those of
the full sample.

The results for the higher mass bin, containing about $\sim 10\%$ of
the galaxies, are shown in the third column of
\autoref{fig:rbfrac}. The third column of \autoref{fig:rbfrac} shows
that even for the galaxies in the higher mass bin at $z=3$, we observe
a higher blue fraction and a lower red fraction in clusters compared
to filaments and sheets. However, the red fraction steadily increases
across all environments as redshift decreases, indicating that star
formation is gradually suppressed in massive galaxies regardless of
their environment. Interestingly, we note that massive galaxies in
voids experience an increase in star formation between $z=2$ and
$z=1$, unlike similarly massive galaxies in other environments. The
sharp increase also indicates that massive galaxies in voids may
follow different evolutionary trajectories compared to those in other
cosmic web environments.  The gas from the expanding void regions can
significantly enhance star formation activity in galaxies residing
within them \citep{ceccarelli12, ceccarelli24}. The absence of the
$z=3$ point in the third column of \autoref{fig:rbfrac} is due to the
lack of galaxies with masses between
$10.5<\log(\frac{M_{*}}{M_{\odot}})<12$ in voids at $z=3$.

\subsection{Stellar mass dependence of the red and blue fractions in different cosmic web environments and their evolution}
\label{subsec:rbfracm}
We analyze the red and blue galaxy fractions as functions of stellar
mass across different cosmic web environments, as shown in
Figures~\ref{fig:bfracmass} and~\ref{fig:rfracmass}. At $z = 3$,
clusters exhibit a higher blue fraction and lower red fraction across
most mass ranges. Filaments show intermediate blue fractions that are
lower than clusters but higher than sheets and voids. However, the
larger uncertainties due to a smaller number of galaxies in sheets and
voids reduce the statistical significance of these differences.

From $z = 3$ to $z = 1.5$, the blue fraction rises significantly for
low-mass galaxies ($\log(M_{\star}/M_{\odot}) \leq 10.5$) in all
environments, with minimal environmental dependence. A clear
environment-dependent trend emerges at $z < 1$ for low-mass galaxies,
where the red fraction increases in clusters from $\sim 40\%$ at $z=1$
to $\sim 70\%$ at $z=0$. At $z=0$, $\sim 40\%$ low mass galaxies in
filaments are red compared to $\sim 30\%$ in sheets and voids. In
contrast, for high-mass galaxies ($\log(M_{\star}/M_{\odot}) > 10.5$),
the red fraction increases across all environments from $\sim 80\%$ to
$\sim 95\%$ suggesting that quenching in massive galaxies is largely
independent of environment.

We observe a peak in the blue fraction at $\log(M_{\star}/M_{\odot})
\sim 10.5$ between $z=2$ and $z=1$ across all cosmic web environments
(\autoref{fig:bfracmass}). This indicates that intermediate-mass
galaxies were the most actively star-forming population during this
epoch, regardless of their location within the cosmic web. The
presence of this peak suggests that galaxies in this mass range had
sufficient gas reservoirs and moderate feedback regulation, enabling
sustained star formation before the onset of widespread quenching
mechanisms at lower redshifts. The sharp increase in the red fraction
at $\log(M_{\star}/M_{\odot}) \sim 10.5$ (\autoref{fig:rfracmass})
marks a critical transition in galaxy evolution, where galaxies shift
from being primarily star-forming to increasingly quenched. Below this
mass threshold, environmental factors dominate quenching, while
internal mechanisms like mass quenching and AGN feedback take over
above it \citep{birnboim03, dekel06, weinberger18, donari21, mao22,
  hasan23}.

\subsection{Evolution of relative red and blue fractions in different cosmic web environments}
\label{subsec:rrbfrac}

\begin{figure}
    \centering
    \includegraphics[width = 0.9\textwidth]{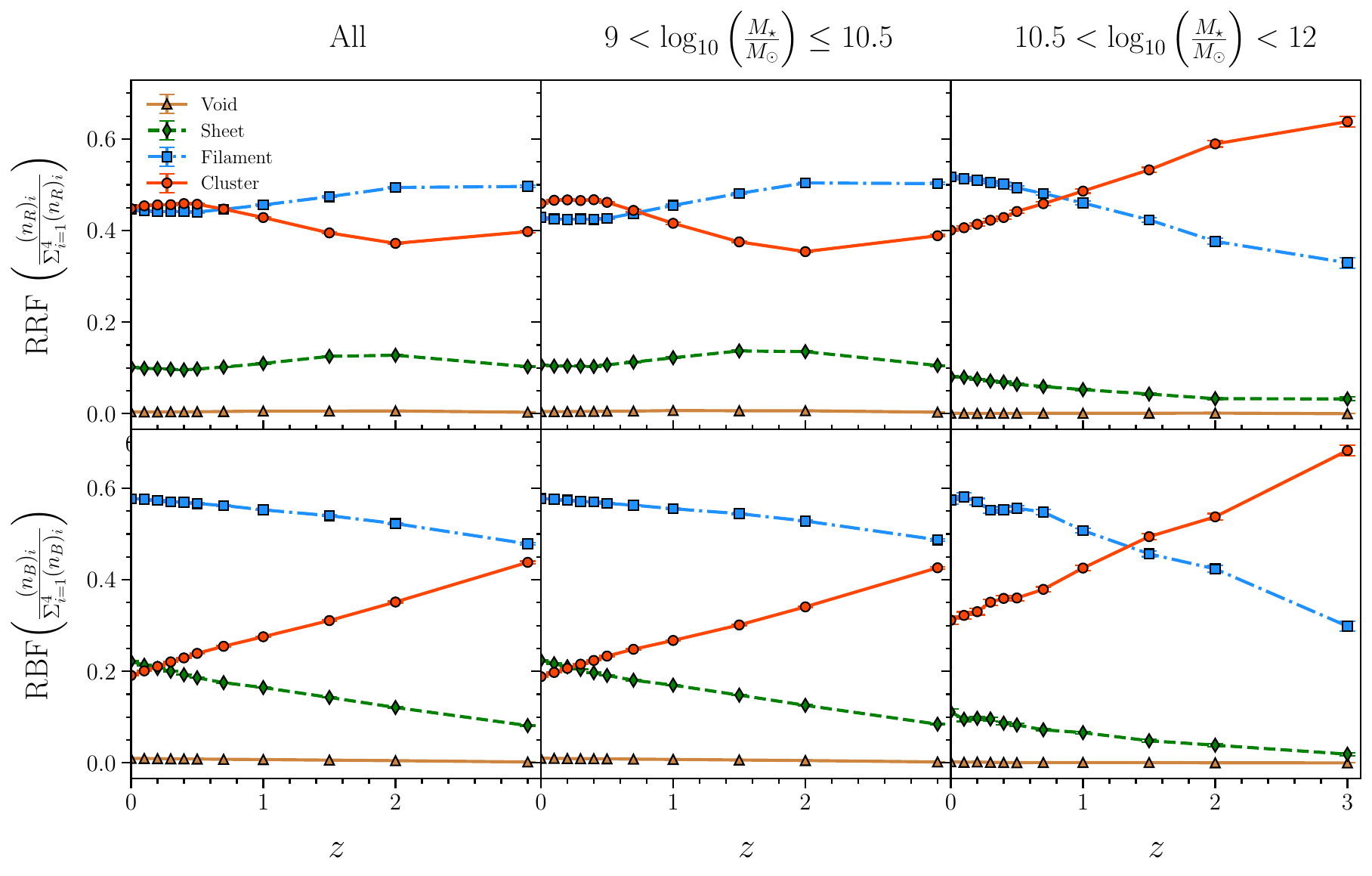}
    \caption{The panels in this figure show the evolution of the
      relative red fraction (RRF) and relative blue fraction (RBF) in
      different cosmic web environments with redshift for different
      mass ranges. The errorbar shown at each data point represents
      the 1$\sigma$ Binomial errorbar.}
    \label{fig:rrbfrac}
\end{figure}

We examine the evolution of the RRF and RBF across different cosmic
web environments, as shown in the top and bottom left panels of
\autoref{fig:rrbfrac}. The top and bottom left panels of
\autoref{fig:rrbfrac} reveal that in filaments and sheets, the RRF
decreases and the RBF increases with decreasing redshift. Conversely,
clusters show an opposite trend where the RRF increases and the RBF
decreases as redshift decreases. Throughout the entire redshift range,
filaments consistently have the highest RBF. At lower redshifts
($z<1$), clusters exhibit the highest RRF compared to other
environments.

The middle and right panels of \autoref{fig:rrbfrac} present the RRF
and RBF for lower and higher mass bins, respectively. The trends
observed in the lower mass bin mirror those of the entire sample,
given that most galaxies belong to the lower mass category. In the
higher mass bin, the behavior of the RRF and RBF is less
intuitive. Specifically, both RRF and RBF decrease in clusters with
decreasing redshift, while they increase in filaments and sheets. For
higher mass galaxies, clusters have higher RRF and RBF than filaments
at earlier times ($z>1$), but this trend reverses at $z\sim 1$, with
filaments surpassing clusters in both RRF and RBF. This
counter-intuitive result can be explained by examining the relative
proportions of galaxies in different cosmic web environments.

The anomalous results in the right panels of \autoref{fig:rrbfrac} can
be explained by the trends observed in the right panel of
\autoref{fig:frac} and \autoref{fig:rbfrac}. For high mass galaxies,
RF increases and BF decreases in all types of cosmic web environments,
with comparable values across different environments
(\autoref{fig:rbfrac}). Consequently, the magnitudes of the RRF and
RBF are primarily determined by the proportions of galaxies in
different environments. The trends observed in the right panel of
\autoref{fig:frac} are reflected in the top and bottom right panels of
\autoref{fig:rrbfrac}, with crossovers in RRF and RBF occurring at
different redshifts due to the varying red and blue fractions in
filaments and clusters.

\subsection{Stellar mass dependence of the relative red and blue fractions in different cosmic web environments and their evolution}
\label{subsec:rrbfracm}

\begin{figure}
    \centering
    \includegraphics[width = 0.9\textwidth]{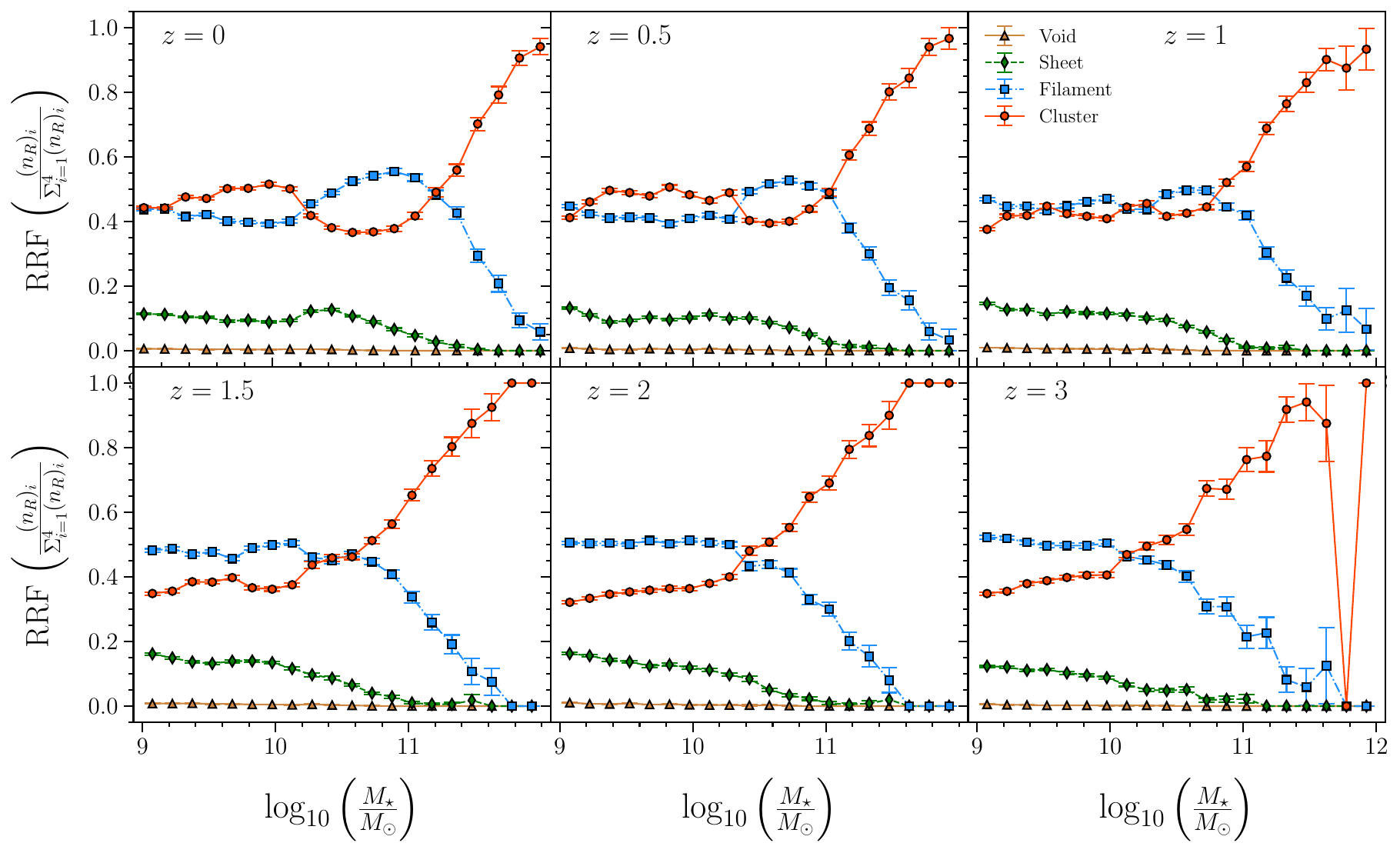}
    \caption{This shows the RRF as a function of stellar mass for
      different geometric environments at different redshifts between
      $3$ to $0$. We show the 1$\sigma$ Binomial errorbars at each
      data point.}
    \label{fig:rrfracmass}
\end{figure}

\begin{figure}
    \centering
    \includegraphics[width = 0.9\textwidth]{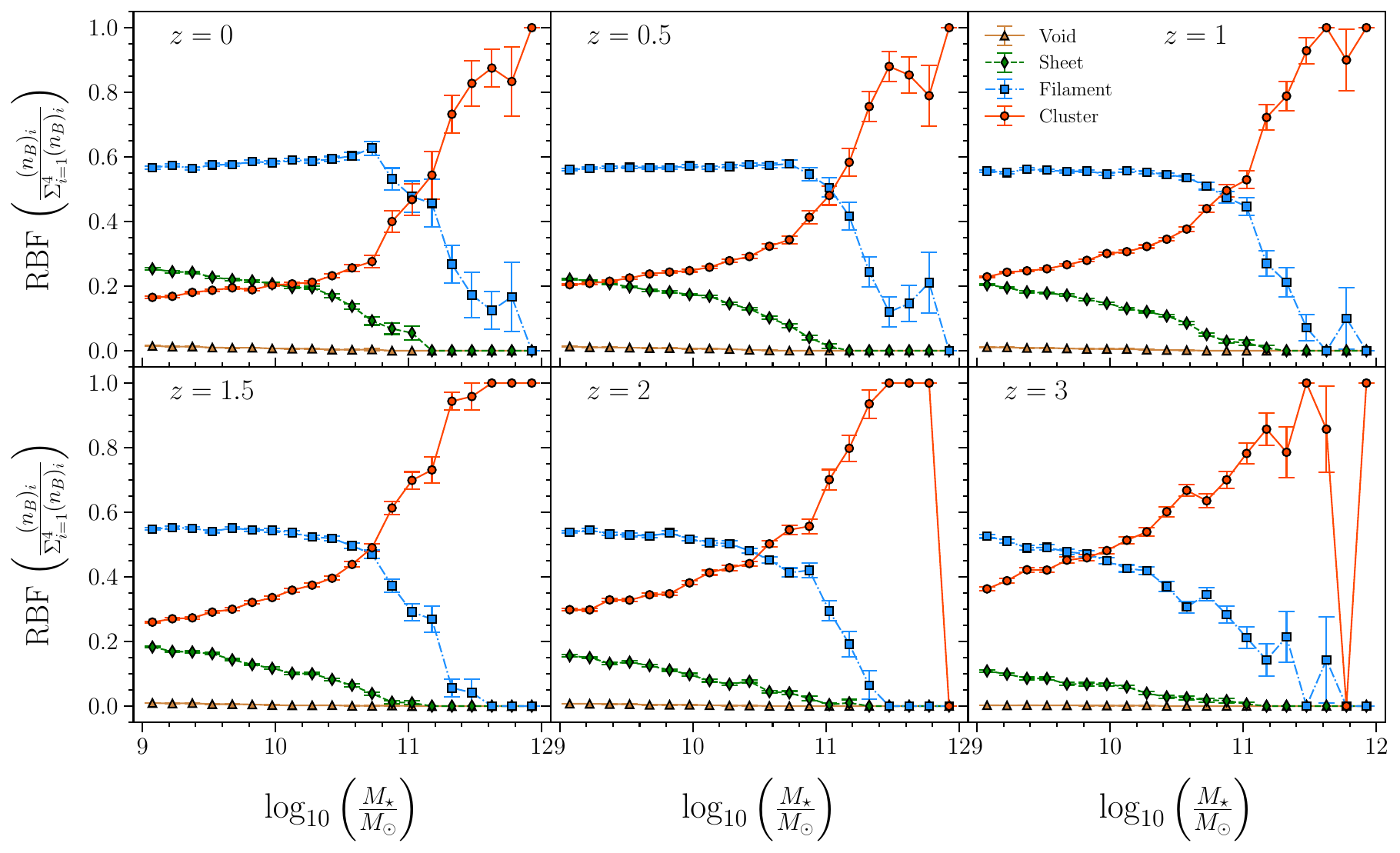}
    \caption{Same as \autoref{fig:rrfracmass} but for RBF.}
    \label{fig:rbfracmass}
\end{figure}

We examine the evolution of the RRF and RBF as functions of stellar
mass and cosmic web environment from $z=3$ to $z=0$
(\autoref{fig:rrfracmass}, \autoref{fig:rbfracmass}). In filaments and
sheets, the RRF is relatively higher at low stellar masses compared to
higher masses and decreases with decreasing redshift, indicating that
low-mass galaxies in these environments tend to remain blue for longer
periods. Conversely, in clusters, the RRF is initially higher at high
masses but increases at low masses over time, suggesting that
environmental quenching of low-mass galaxies becomes more prominent at
late times.

At $z \sim 1$, the RRF at low masses becomes comparable in filaments
and clusters. By $z=0$, however, clusters dominate the red population
at both low and high stellar masses, while filaments show a peak in
RRF at intermediate masses ($10.5 < \log(M_{\star}/M_{\odot}) <
11$). Two notable crossovers in the RRF curves at $z=0$ illustrate
this shift: at low masses, clusters overtake filaments in RRF,
indicating intensified quenching in dense environments; and at
intermediate masses, filaments surpass clusters, emerging as the
primary hosts of quiescent galaxies in that mass range.

RBF trends provide a complementary view. Clusters consistently show
lower RBF values at low stellar masses across redshifts, which is
consistent with clusters hosting lower fractions of blue, low-mass
galaxies. In contrast, filaments show the highest RBF at intermediate
and low masses for $z < 1$, highlighting their transitional role
hosting both star-forming and quenched galaxies. This contrasts with
clusters, which are associated with efficient quenching and
consistently exhibit low RBF values compared to filaments across
redshifts. Voids, on the other hand, contain a very small number of
galaxies overall, resulting in low RRF and RBF values at all stellar
masses and redshifts. While their low-density nature suggests a
star-forming preference, the small sample size limits statistical
interpretation.

These trends support the idea that mass-driven and environment-driven
quenching processes operate in tandem, but on different timescales and
with varying efficiencies across environments. Filaments appear to
quench galaxies earlier, particularly at low stellar masses, while
clusters become increasingly effective at quenching low-mass galaxies
at later times. For intermediate-mass galaxies, filaments play a
dominant role significantly influencing their evolutionary
pathways. Our results are consistent with previous findings
\citep{salerno19, mao22}, which show that low-mass galaxies in dense
environments quench rapidly, whereas intermediate-mass galaxies are
more sensitive to the surrounding large-scale
structure. Interestingly, clusters also exhibit the highest RBF at the
high-mass end ($\log(M_{\star}/M_{\odot}) > 11$) even at low redshift,
possibly reflecting episodes of renewed star formation triggered by
mergers or dynamical interactions in these dense environments.

\begin{figure}[htbp!]
\centering
\includegraphics[width = 0.9\textwidth]{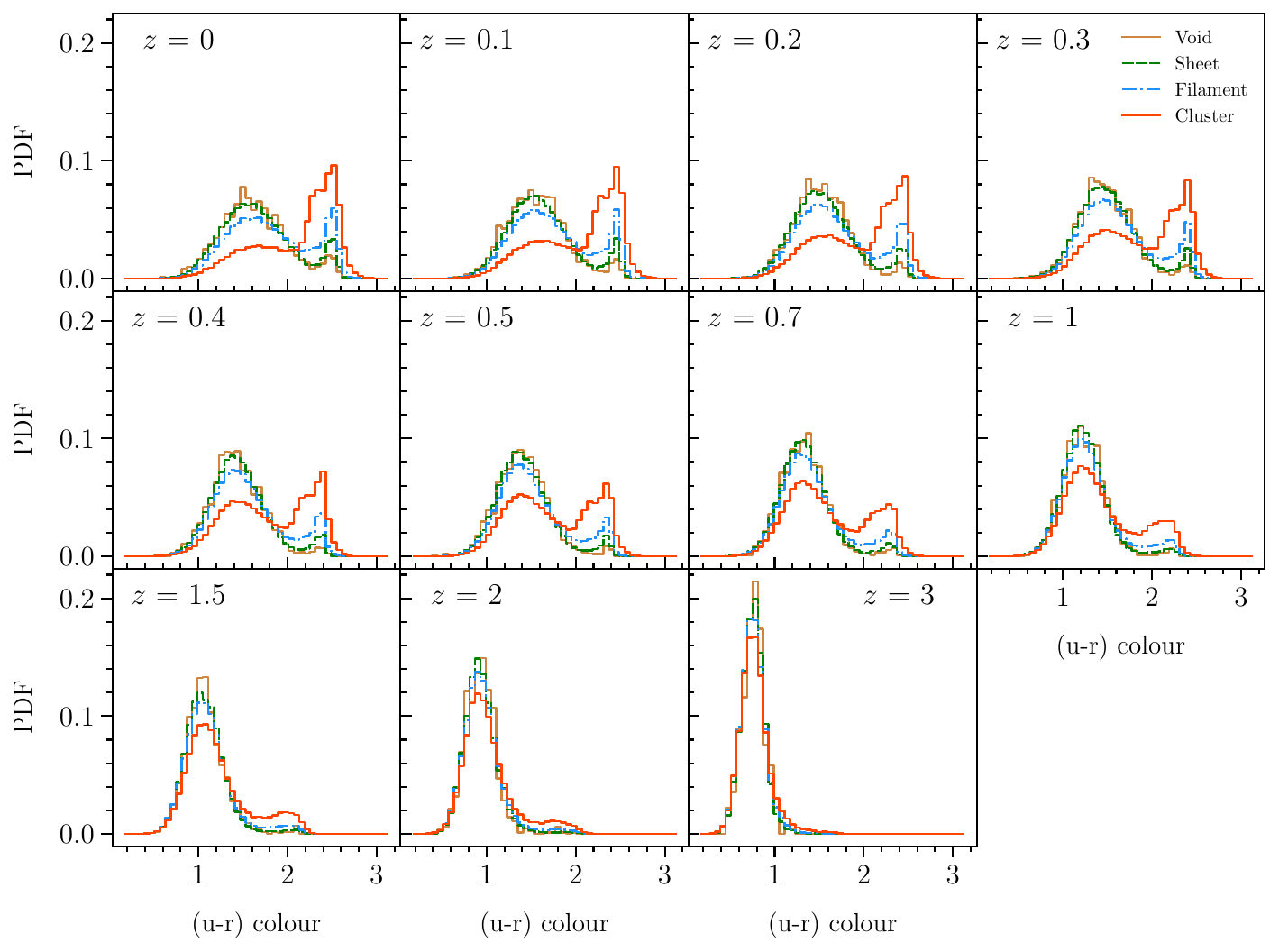}
\caption{This figure shows the PDF of $(u-r)$ colour in different
  cosmic web environments at different redshifts.}
\label{fig:colpdf}
\end{figure}

\begin{figure}[htbp!]
\centering
\includegraphics[width = 0.4\textwidth]{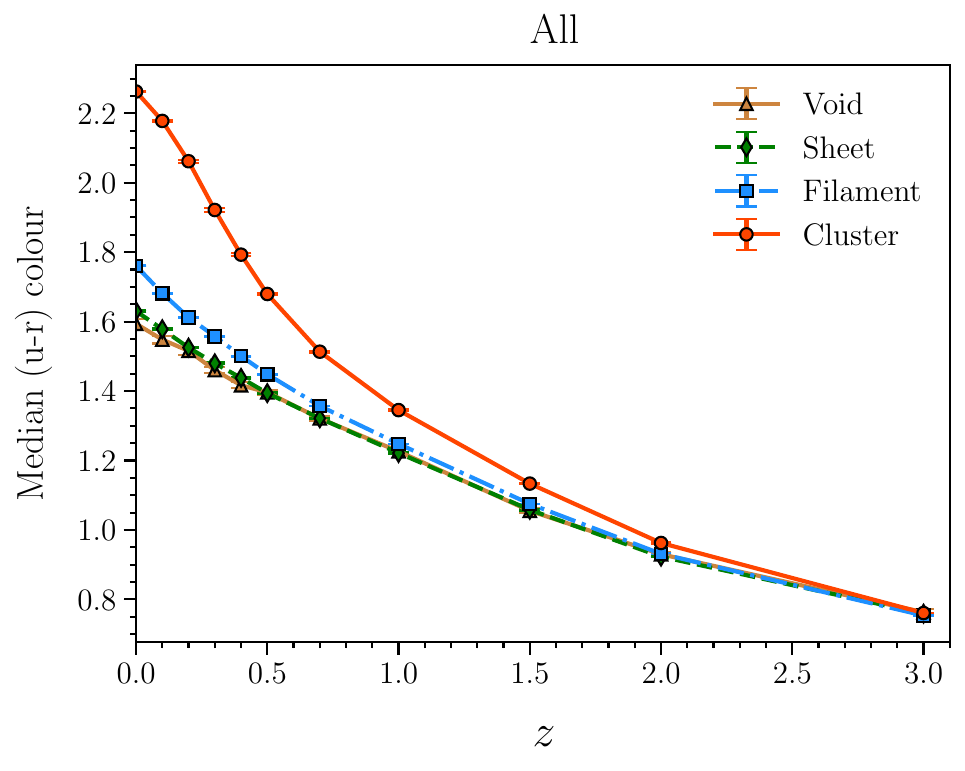}
\includegraphics[width = 0.4\textwidth]{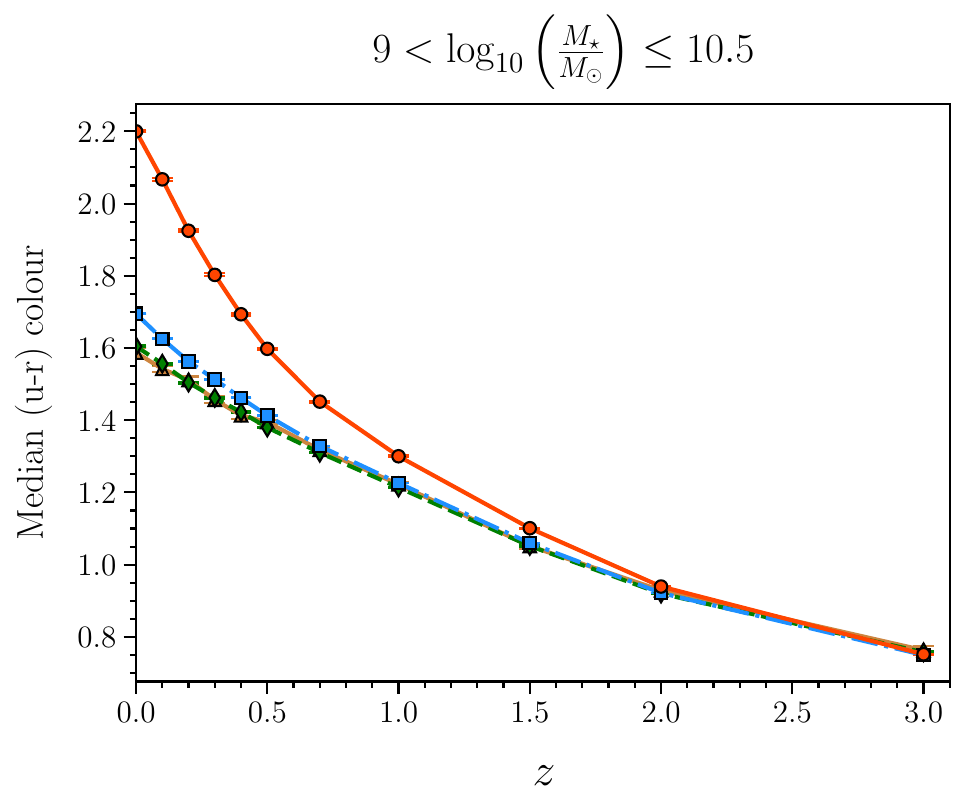}
\includegraphics[width = 0.4\textwidth]{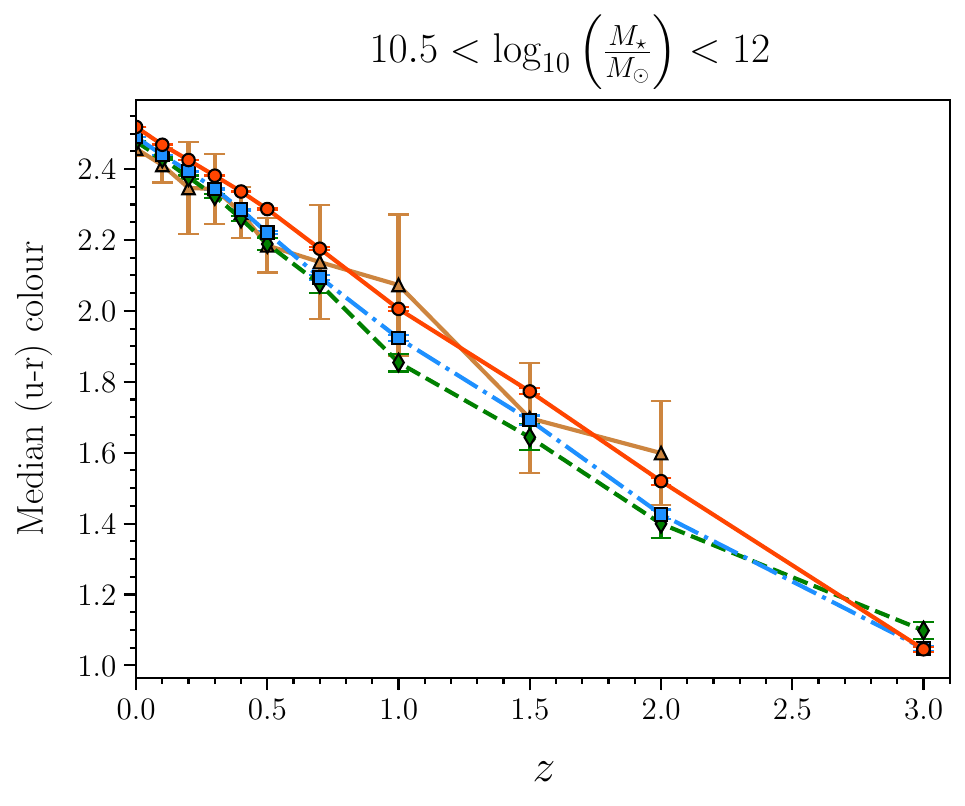}
\caption{This figure shows the evolution of median $(u-r)$ colour in
  different cosmic web environments with redshift. The top left panel
  shows the results for the entire sample. The top right panel and the
  bottom panel show the results for the lower and higher mass bin
  respectively. The 1$\sigma$ errorbars are calculated using $50$
  bootstrapped realizations from the original data.}
\label{fig:colmed}
\end{figure}

The trends observed in all the galaxy fractions across cosmic web
environments may, in part, be driven by variations in local density,
since denser regions naturally host different types of galaxies
compared to underdense areas. To disentangle the effects of local
density from those of the geometric environment, it is essential to
examine the fractions F, RF, BF, RRF, and RBF within narrow density
bins. This approach would allow us to assess whether the observed
environmental trends are intrinsic to the cosmic web environment or
simply a reflection of underlying density variations, providing a
clearer picture of the role played by cosmic web in galaxy
evolution. We address this issue in detail in \autoref{subsec:locden}
by analyzing the galaxy fractions within narrow density bins to
isolate the influence of cosmic web geometry from that of density.

\subsection{Evolution of colour across different cosmic web environments}
\label{subsec:col}

We present the distributions of $(u-r)$ colour across various cosmic
web environments at different redshifts, ranging from $z=3$ to $z=0$,
as shown in different panels of \autoref{fig:colpdf}. At $z=3$, the
$(u-r)$ colour distribution in each environment is unimodal, with a
peak around $(u-r) \sim 0.7$. As redshift decreases, this peak shifts
towards higher $(u-r)$ values, indicating the aging of stellar
populations. By $z=2$, the colour distribution begins to show
bimodality. The bimodality becomes most pronounced in clusters,
followed by filaments, sheets, and voids. In clusters, the peak
corresponding to the red sequence grows rapidly, eventually surpassing
the peak of the blue cloud in this environment at $z=0.5$. The
emergence of distinct bimodality across all environments suggests that
the colour transformation can occur in any cosmic web environment.

In the top left panel of \autoref{fig:colmed}, we show the median
$(u-r)$ colour of galaxies in various cosmic web environments as a
function of redshift. Since $z \sim 2$, the median colour in clusters
begins to diverge from that in other environments and increases
rapidly after $z > 1$. Meanwhile, the median colour in filaments
starts to differ from that in sheets and voids after redshift 1. At $z
< 1$, clusters exhibit the highest median colour, followed by
filaments, sheets, and voids.

The top right and bottom panels of \autoref{fig:colmed} illustrate the
evolution of the median colour in each environment for lower and
higher mass bins. The lower mass bin shows trends similar to the
combined sample due to the large number of galaxies in this bin. The
bottom panel reveals that the median colour for more massive galaxies
($10.5 < \log(\frac{M_{*}}{M_{\odot}}) < 12$) increases steadily at $z
< 1$, with little variation across different environments. This
suggests that the colour evolution of the more massive galaxies is
primarily driven by mass and is less dependent on their
environment. This reaffirms our findings in \autoref{subsec:rbfrac}
and \autoref{subsec:rbfracm}.

\begin{figure}[htbp!]
\centering
\includegraphics[width = 0.9\textwidth]{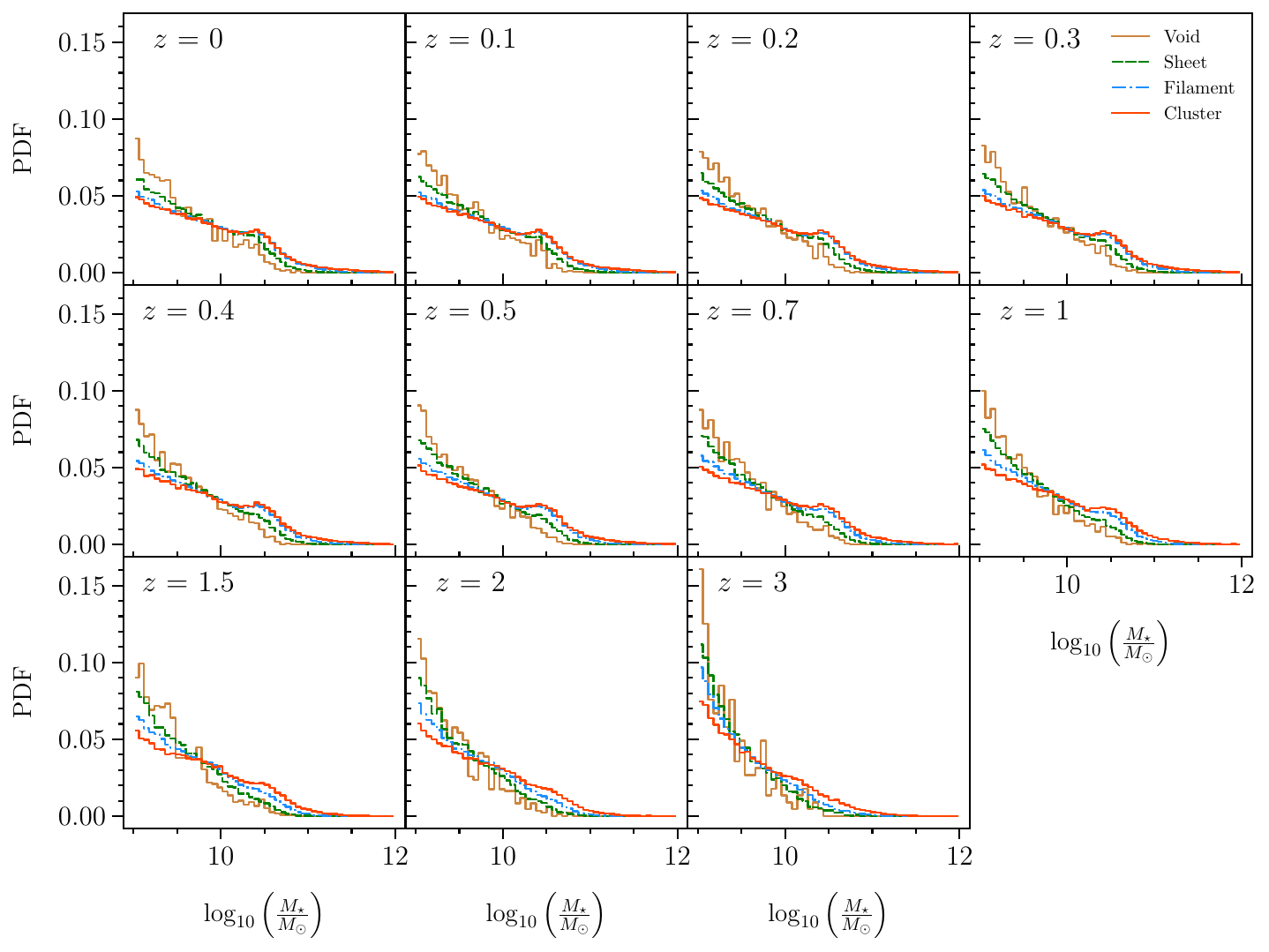}
\caption{Same as \autoref{fig:colpdf} but for stellar mass.}
\label{fig:masspdf}
\end{figure}

\begin{figure}[htbp!]
\centering
\includegraphics[width = 0.4\textwidth]{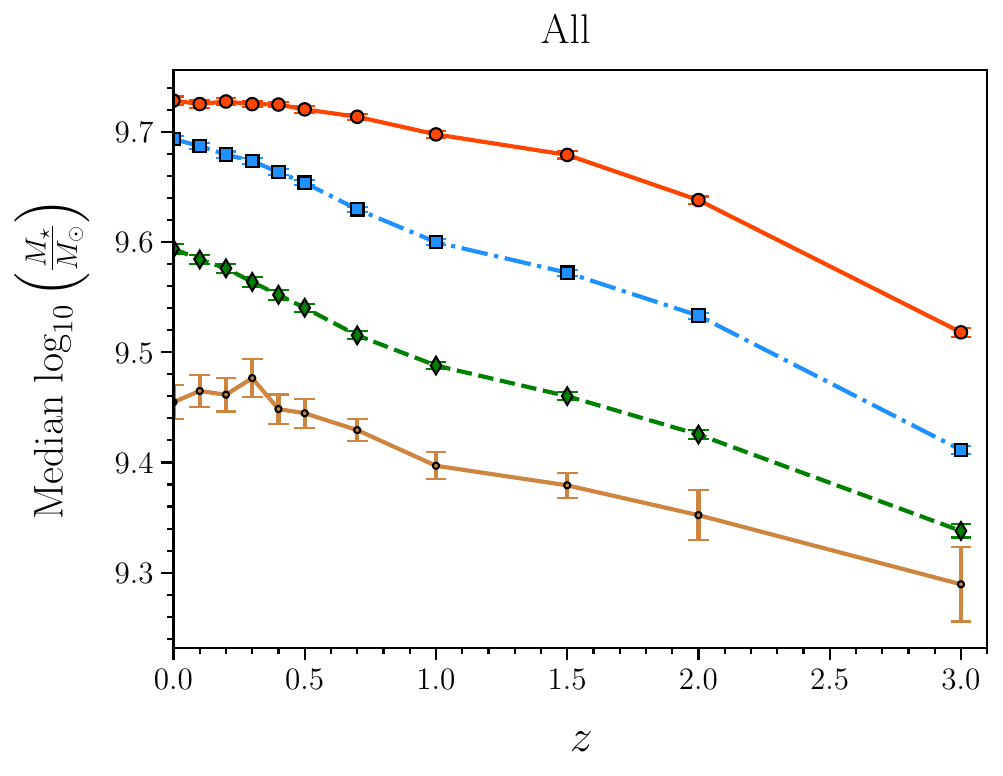}
\includegraphics[width = 0.4\textwidth]{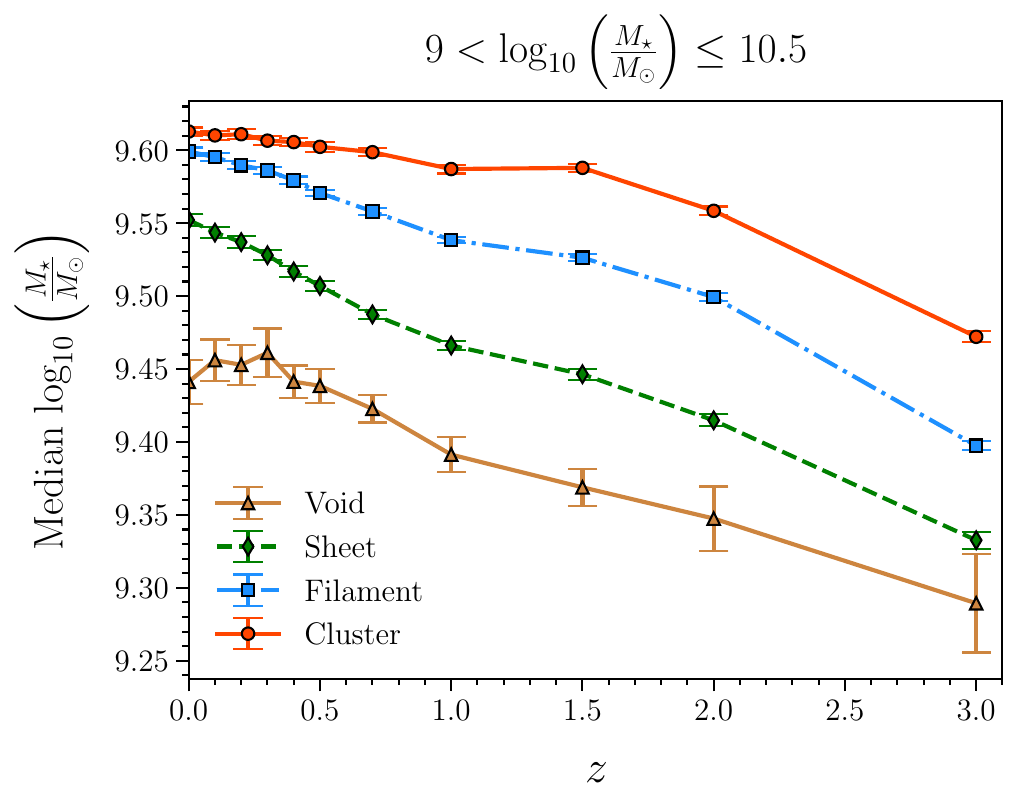}
\includegraphics[width = 0.4\textwidth]{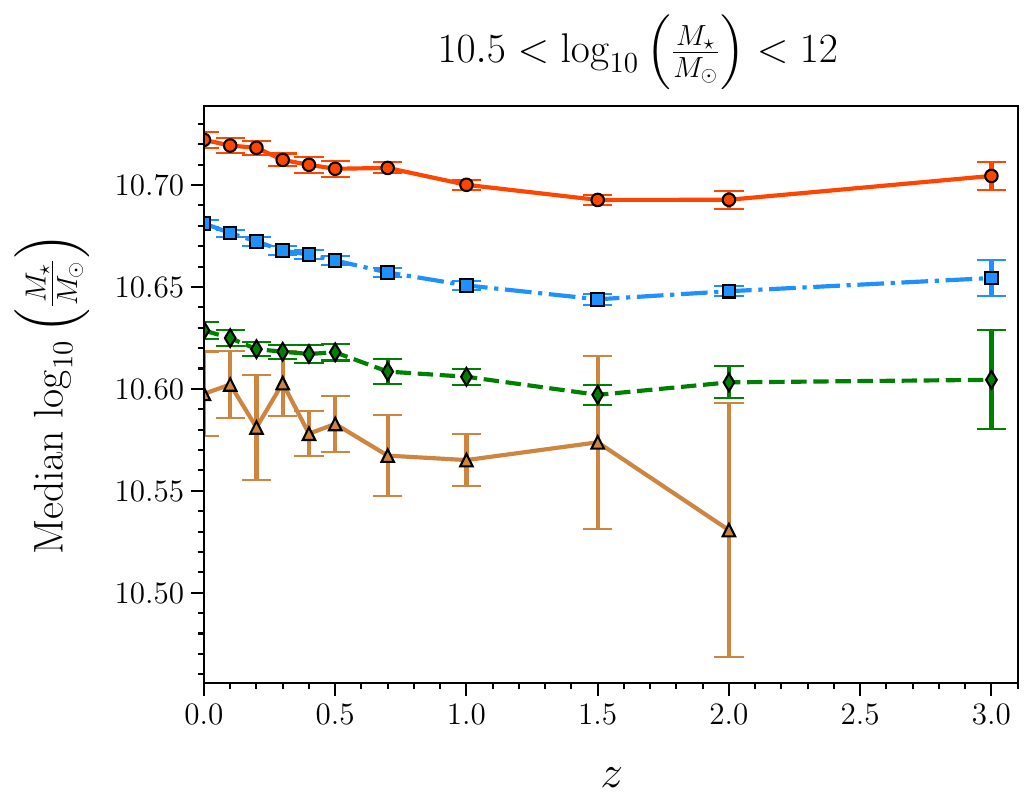}
\caption{Same as \autoref{fig:colmed} but for stellar mass.}
\label{fig:massmed}
\end{figure}

\subsection{Evolution of stellar mass across diverse cosmic web environments}
\label{subsec:mass}

We present the stellar mass distributions of galaxies across different
environments from $z=3$ to $z=0$ in \autoref{fig:masspdf}. At higher
masses, galaxies are predominantly found in clusters, followed by
filaments, sheets, and voids. Conversely, at lower masses, the
distribution is dominated by galaxies in voids, with fewer in sheets,
filaments, and clusters. This trend indicates that massive galaxies
are more common in clusters and filaments, while less massive galaxies
are more prevalent in sheets and voids. At $z<2$, we observe the
emergence of a bump in the stellar mass distribution around
$\log(\frac{M_{*}}{M_{\odot}} \sim 10.5$) in each environment. The
bump becomes more pronounced in all environments after $z=1$
suggesting an increase in the number of intermediate-mass galaxies
during this period.

In the top left panel of \autoref{fig:massmed}, we show the evolution
of the median stellar mass of galaxies across different cosmic web
environments. It shows that as redshift decreases, the median mass of
galaxies increases in all environments. At each redshift, the median
mass is highest in clusters, followed by filaments, sheets, and
voids. The top right and bottom panels of \autoref{fig:massmed} show
the results for lower and higher mass bins, respectively. In the top
right panel, we observe that the median mass of galaxies in filaments
and sheets increases steadily between $z=3$ to $z=1$. However, at
$z<1$, the growth rate of the median mass for lower mass galaxies in
clusters slows down significantly compared to those in filaments and
sheets. The bottom panel shows that the median stellar mass for high
mass galaxies remains stable between $z=3$ and $z=1.5$, followed by a
gradual increase up to the present. This suggests that these galaxies
undergo a period of minimal mass growth during the earlier stages of
cosmic evolution, likely indicating that major mergers are less
frequent at higher redshifts. The steady increase in median stellar
mass at $z<1.5$ points to more significant mass accumulation in later
stages, possibly due to mergers and accretion.

It may be interesting to note that while the median colour of massive
galaxies is quite similar across different environments (as shown in
\autoref{fig:colmed}), their median mass differs significantly. 

\begin{figure}[htbp!]
\centering
\includegraphics[width = 0.9\textwidth]{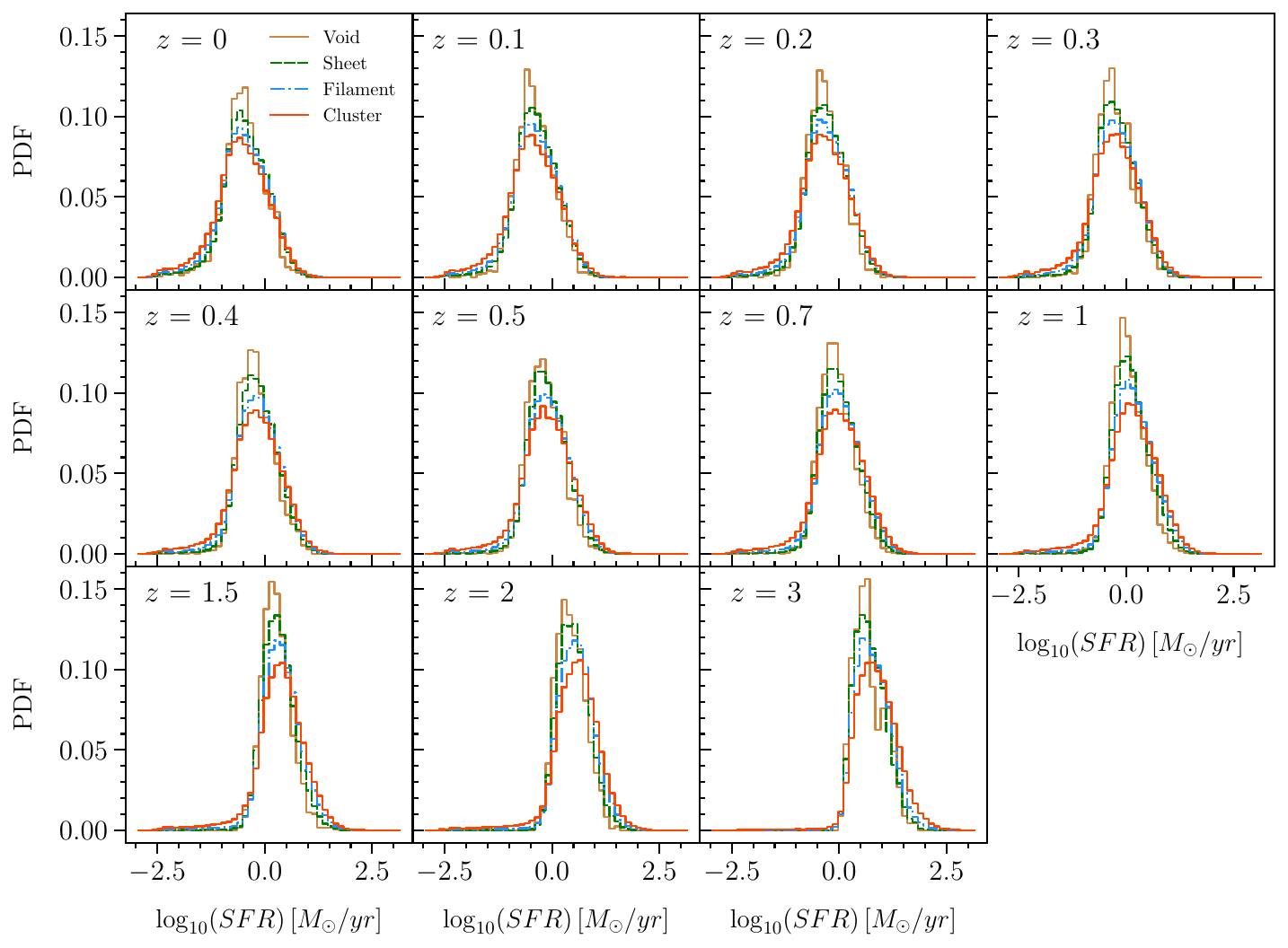}
\caption{Same as \autoref{fig:colpdf} but for SFR.}
\label{fig:sfrpdf}
\end{figure}

\begin{figure}[htbp!]
\centering
\includegraphics[width = 0.4\textwidth]{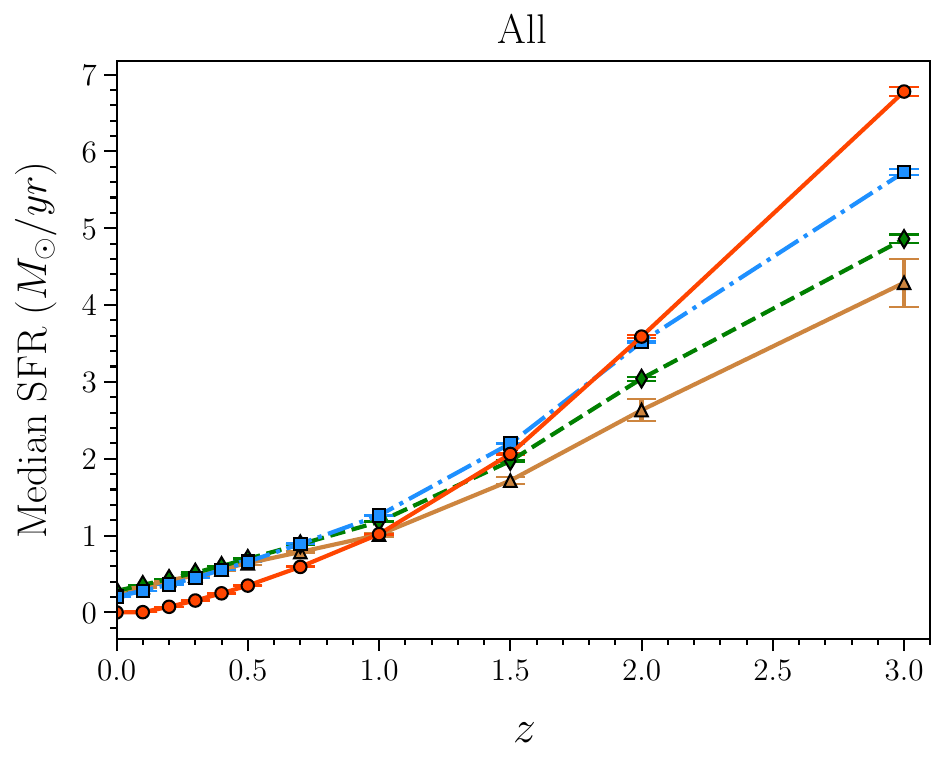}
\includegraphics[width = 0.4\textwidth]{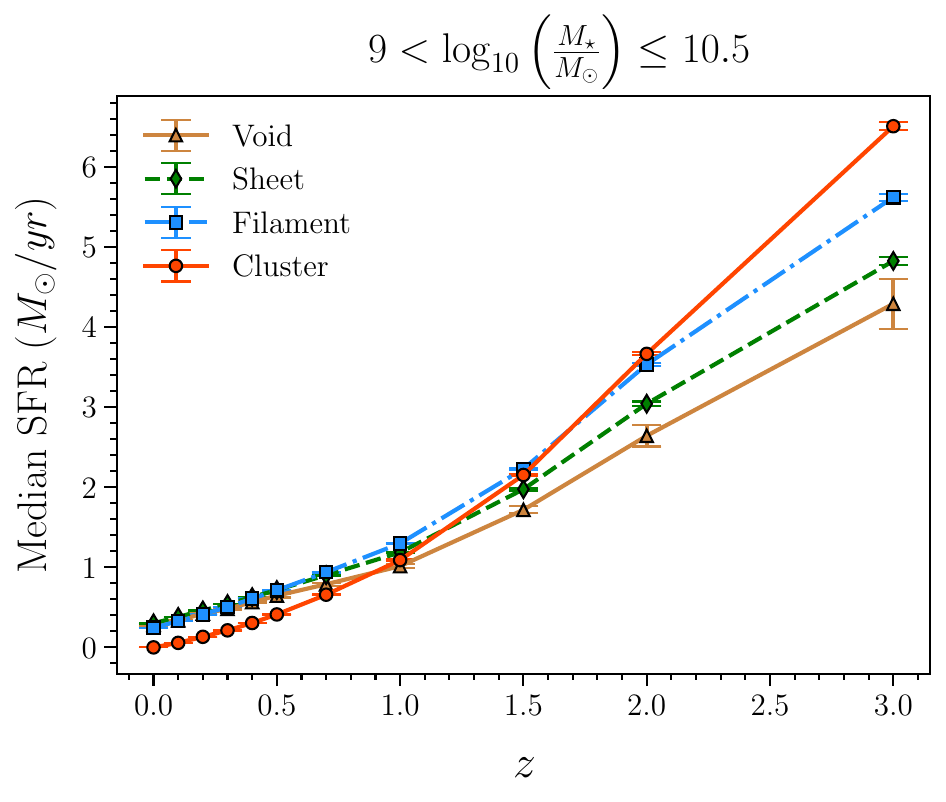}
\includegraphics[width = 0.4\textwidth]{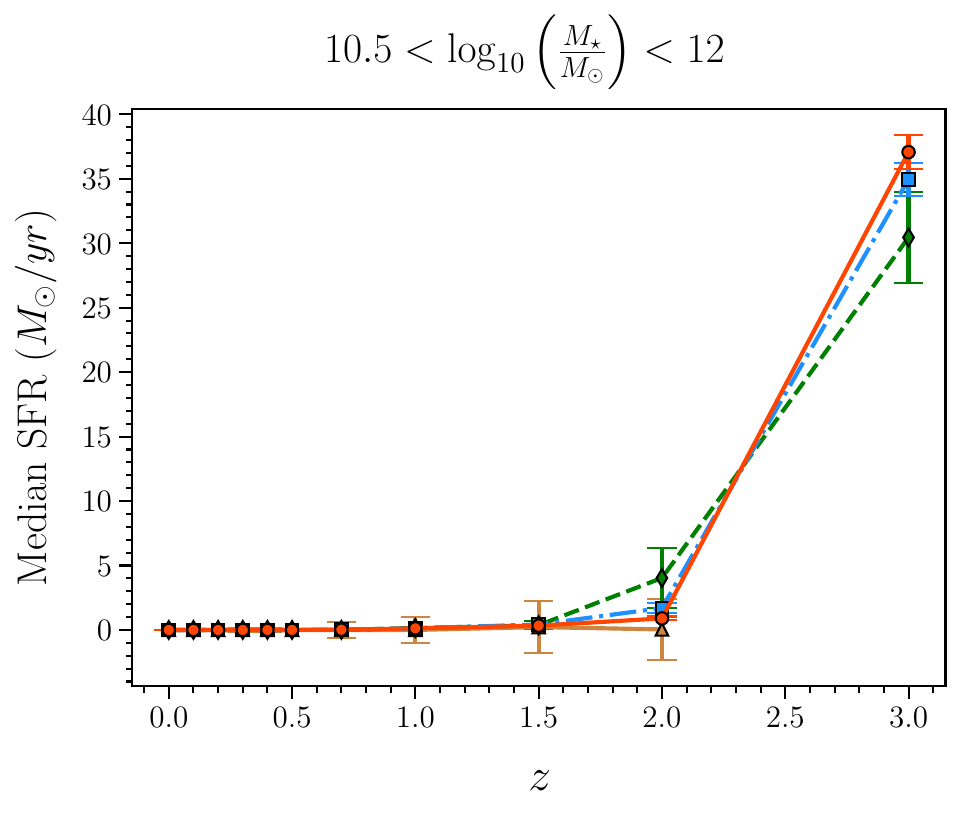}
\caption{Same as \autoref{fig:colmed} but for SFR.}
\label{fig:sfrmed}
\end{figure}

\begin{figure}[htbp!]
\centering
\includegraphics[width = 0.9\textwidth]{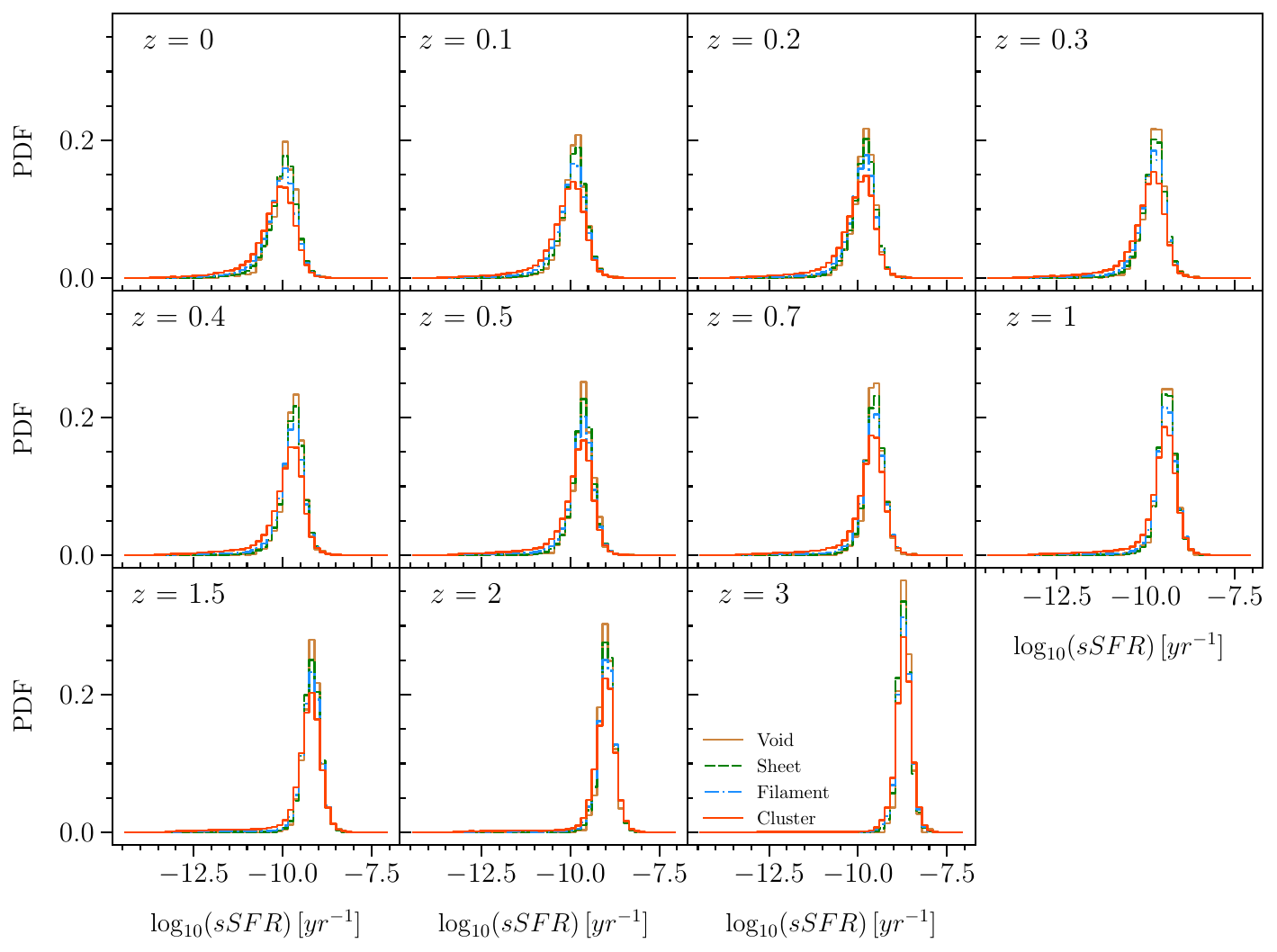}
\caption{Same as \autoref{fig:colpdf} but for sSFR.}
\label{fig:ssfrpdf}
\end{figure}

\begin{figure}[htbp!]
\centering
\includegraphics[width = 0.4\textwidth]{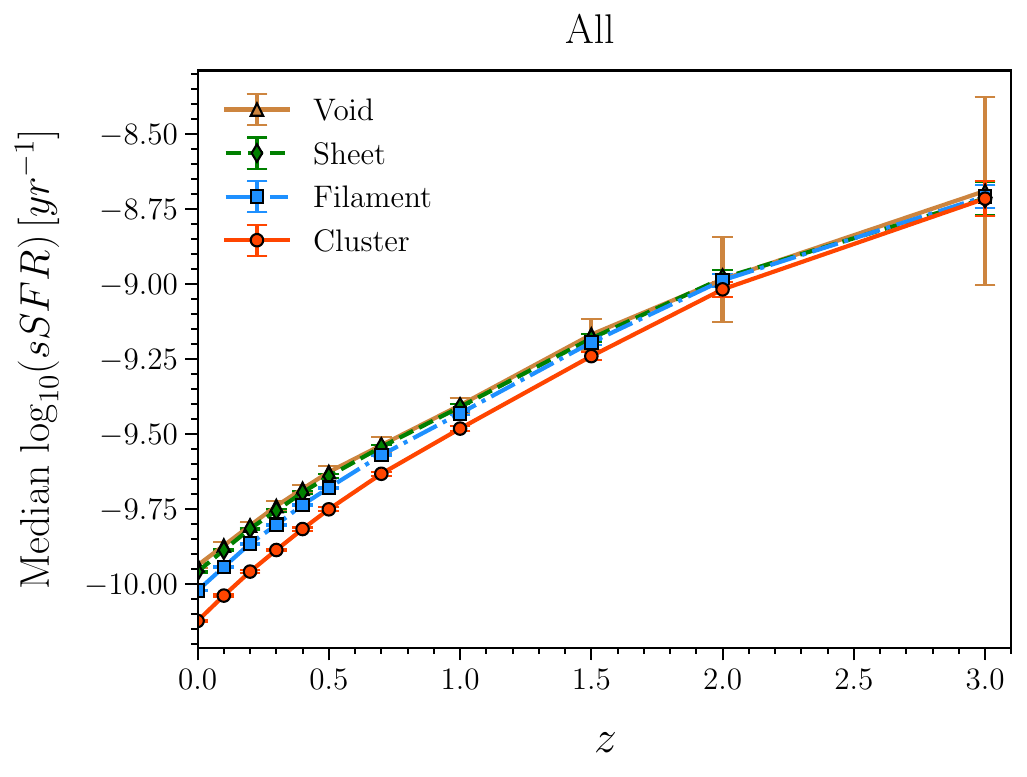}
\includegraphics[width = 0.4\textwidth]{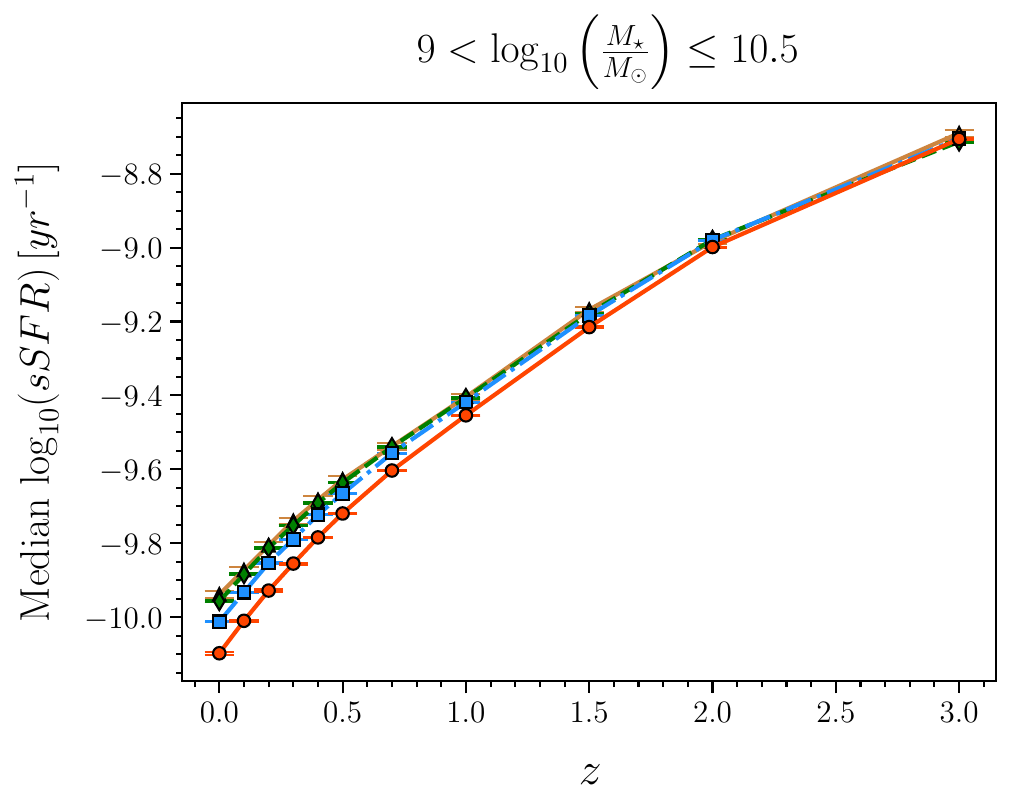}
\includegraphics[width = 0.4\textwidth]{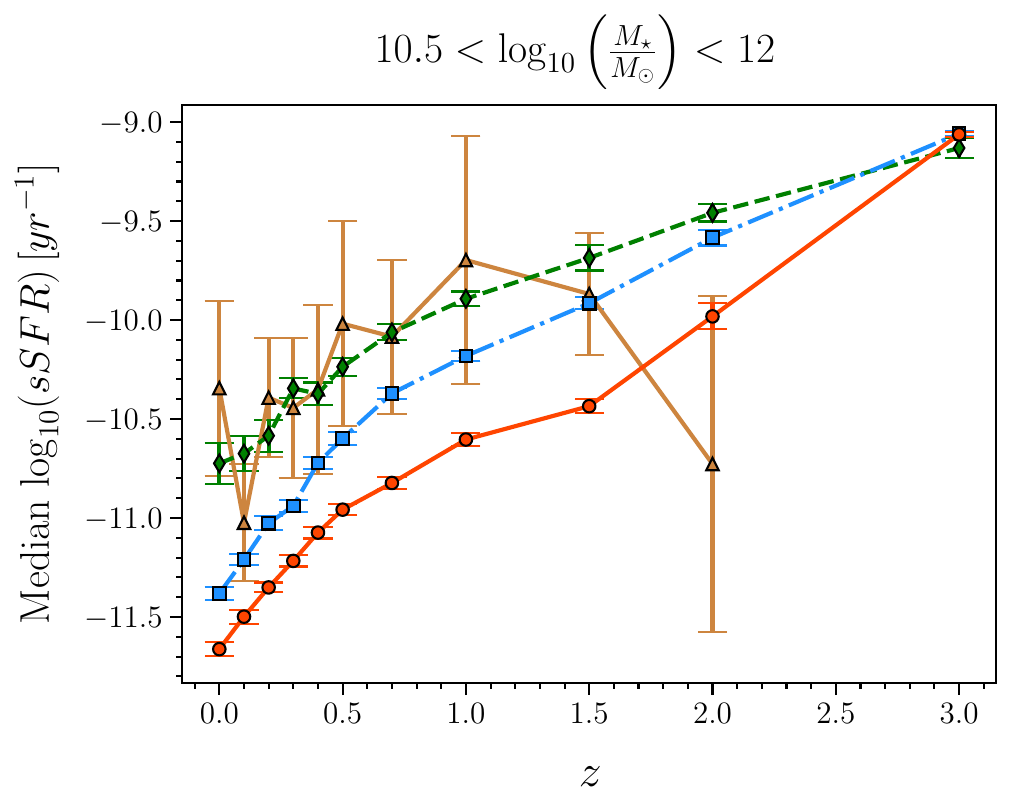}
\caption{Same as \autoref{fig:colmed} but for sSFR.}
\label{fig:ssfrmed}
\end{figure}

\subsection{Evolution of SFR across diverse cosmic web environments}
\label{subsec:sfr}

The SFR distributions of galaxies in various cosmic web environments
are shown at different redshifts in the panels of
\autoref{fig:sfrpdf}. The peak location of the SFR distribution shifts
from $\log(SFR) = 0.75$ to $\log(SFR) = -0.5$ over the redshift range
$z = 3$ to $z = 0$. While the peak locations differ slightly across
cosmic web environments between redshifts $z = 3$ and $z = 1$, they
converge to nearly the same $\log(SFR)$ value at $z<1$. The peak
amplitude is highest in voids, followed by sheets, filaments, and
clusters.

We show the evolution of the median SFR in different environments in
the three panels of \autoref{fig:sfrmed}. The top left panel shows
that the median SFR is highest in clusters, followed by filaments,
sheets, and voids until $z = 1.5$. The median SFR in clusters becomes
smaller than the other geometric environments at $z<1$. The results
for the lower mass bin, shown in the top right panel of
\autoref{fig:sfrmed}, exhibit a similar trend to that observed in the
entire sample. The bottom panel of \autoref{fig:sfrmed} indicates that
the median SFR for massive galaxies ($\log(\frac{M_{*}}{M_{\odot}}) >
10.5$) drops to zero at $z \sim 1.5$ across all environments,
suggesting that star formation in high-mass galaxies is strongly
suppressed in the low-redshift universe.

\subsection{Evolution of sSFR across diverse cosmic web environments}
\label{subsec:ssfr}

We show the distribution of specific star formation rate (sSFR) across
cosmic web environments at various redshifts in the panels of
\autoref{fig:ssfrpdf}. While the star formation rate (SFR) measures
the degree of star formation in a galaxy, sSFR normalizes this rate by
its stellar mass. This normalization is crucial for comparing star
formation activity across galaxies of different sizes, offering
insights into the efficiency of star formation relative to galaxy
mass. The panels in \autoref{fig:ssfrpdf} reveal trends similar to
those seen in \autoref{fig:sfrpdf}. Specifically, the peak of the sSFR
distribution in each environment shifts towards more negative values
as redshift decreases. This gradual shift points to both internal and
external mechanisms leading to the aging and transformation of
galaxies in each type of environment. The peak amplitude is highest in
voids, followed by sheets, filaments, and clusters. This indicates
that, for a given stellar mass, star formation is strongest in less
dense environments and becomes progressively weaker as the environment
grows denser.

The median sSFR across different environments is shown in
\autoref{fig:ssfrmed}. The top left panel indicates that, unlike the
median SFR, there are no statistically significant differences in the
median sSFR across environments in the redshift range $1.5 < z <
3$. As redshift decreases, sSFR decreases in all environments. At $z <
1$, clusters exhibit the lowest median sSFR, with filaments showing
somewhat higher sSFR compared to clusters between $z = 1$ and $z =
0$. Sheets and voids have the highest and similar median sSFR in this
redshift range. The top right panel presents results for the lower
mass bin, which mirrors the trends observed in the entire sample for
the reasons discussed earlier. The bottom panel of
\autoref{fig:ssfrmed} displays results for the higher mass bin. This
panel shows similar trends but with larger differences in median sSFR
across environments. It is interesting to note that the median sSFR
for massive galaxies in voids tends to increase between $z = 2$ and $z
= 1$, though this trend is difficult to confirm due to large errorbars
for the galaxies in voids.


\section{Discussions and Conclusions}
We study the evolution of red and blue galaxies in different cosmic
web environments using IllustrisTNG simulation. Our key findings are
summarized as follows.

 (i) The evolution of red and blue fractions show strong environmental
dependence. At $z=3$, clusters host the highest fraction of blue
galaxies, but this trend reverses at lower redshifts as environmental
quenching becomes dominant. The inflection points at $z=1.5$ for
clusters, $z=0.7$ for filaments and sheets, and $z=0.5$ for voids mark
key transitions where red galaxies start to dominate in these
environments.

(ii) The environmental quenching plays a major role at low masses. For
low-mass galaxies ($\log(\frac{M_{*}}{M_{\odot}})<10.5$), clusters
host the highest red fractions at $z < 1$, showing that environmental
effects drive quenching in these systems. Galaxies in filaments
exhibit intermediate quenching signatures, spanning a broader range of
evolutionary stages. At $z<1$, low-mass galaxies in filaments show a
higher blue fraction compared to those in clusters.

(iii) The filaments serve as a bridge between the galaxy evolution in
cluster and void type environments. At $z < 1$, filaments host a mix
of quiescent and star-forming galaxies, with nearly $60\%$ of low-mass
galaxies remaining blue. The presence of mixed galaxy populations in
filaments may reflect both the less extreme environmental conditions,
potentially allowing more gradual quenching, and the fact that
filaments host a large fraction of the galaxy population, thus
capturing a broad diversity of evolutionary stages.

(iv) Our findings support the interpretation that stellar mass plays a
significant role in the quenching of high-mass galaxies. For massive
galaxies ($\log(\frac{M_{*}}{M_{\odot}}) > 10.5$), quenching occurs
independent of environment, suggesting that internal processes like
mass quenching and AGN feedback dominate over external effects.

(v) The sharp increase in the red fraction at
$\log(\frac{M_{*}}{M_{\odot}}) \sim 10.5$ marks a transition in galaxy
evolution. It corresponds to a buildup of intermediate-mass galaxies
and represents a shift from star-forming to quiescent populations,
becoming more prominent at $z < 1$.

(vi) The colour distribution transitions from unimodal ($z > 2$) to
bimodal ($z < 2$), with clusters developing the strongest red
peak. The strengthening of colour bimodality over time indicates the
build-up of quiescent galaxies at lower redshifts.

(vii) The median colours of massive galaxies do not differ
significantly across various environments implying that colour
evolution in high mass galaxies is largely driven by mass rather than
environment. The median mass of lower mass galaxies in clusters
exhibit a slower growth at $z<1$. This indicates that environmental
factors influence galaxy mergers or accretion more strongly than
colour evolution.

Our findings align well with previous observational and
simulation-based studies on galaxy evolution within the cosmic web,
each probing specific redshift regimes and scales. For example,
studies such as \cite{kuutma17} (SDSS, $0.02\leq z \leq0.155$),
\cite{malavasi17} (VIPERS, $z\sim0.7$), \cite{laigle18} (COSMOS,
$0.5<z<0.9$), \cite{kraljic18} (GAMA, $0.03\leq z\leq0.25$),
\cite{bonjean20} (WISExSCOS, $0.1<z<0.3$), \cite{welker20} (SAMI,
$0.004<z<0.13$), and \cite{gouin20} (WISE x SCOSMOS, $0.1<z<0.3$)
report that galaxies near filament spines tend to be redder, more
massive, and less star-forming at fixed local density on scales of a
few to several Mpc. However, as pointed out by \cite{okane24} (SDSS,
$z\leq0.077$), some of these trends may be fully explained by local
density effects alone, without invoking additional influence from the
filament geometry.

One widely accepted interpretation is that galaxies migrating toward
filament spines experience elevated accretion rates, more frequent
interactions, and increased merger activity leading to accelerated
mass growth and quenching. This interpretation is supported by
simulation-based work across various redshifts and scales, including
IllustrisTNG and zoom-in hydrodynamical studies \citep{singh20,
  song21, malavasi22, kotecha22, kuchner22, jhee22,
  rowntree24}. Additional processes, such as pre-processing in small
galaxy groups embedded within filaments \citep{sarron19, santiago20}
($z=0.15-0.7$), and potential shock heating in filaments suppressing
star formation especially in satellite galaxies are also implicated by
recent analyses using SIMBA and IllustrisTNG simulations
\citep{bulichi24, hasan24} ($z=0-2$). Our observation of an elevated
relative red fraction in filaments at intermediate stellar masses may
be connected to these combined mechanisms.

We also reinforce the picture of filaments as transitional
environments. Observational studies such as \cite{kraljic20} and
\cite{galarraga23}, spanning $0.5\lesssim z \lesssim 2$, show that
low-mass galaxies in filaments maintain higher star formation rates
than their cluster counterparts at fixed stellar mass and local
density. High-resolution zoom-in simulations by \cite{liao19} find
that anisotropic gas accretion onto filament halos at high redshifts
($z=2.5,4$) enhances their baryon and stellar fractions relative to
field analogs. Dark matter-only and hydrodynamical simulations
\citep{veena18, veena19} ($z=0-3$) similarly indicate that low-mass
halos preferentially accrete perpendicularly to their host filament,
while higher-mass halos accrete along the filament axis. Observations
of HI content in SIMBA simulations \citep{dave19} ($z=0-6$) and
related anisotropic inflow signatures \citep{kraljic20b} ($z=0-2$)
further reinforce this scenario.

Finally, the transition in colour distribution from unimodal at $z>2$
to bimodal at $z<1$ supports findings from the EAGLE simulation
\citep{trayford16, wright19}, which identified buildup of the red
sequence around $z\sim1$ due to AGN feedback and satellite
quenching. The increasing red fractions at
$\log(M_{*}/M_{\odot})>10.5$ are also consistent with the
mass-dependent quenching trends reported in \cite{mao22, taylor23,
  hasan23}. Overall, our results align with the broader consensus that
environment-driven quenching dominates for low-mass galaxies, while
mass- and AGN-driven quenching become increasingly important at higher
masses \citep{weinberger18, donari21}. Throughout our analysis, the
cosmic web classification is based on a Gaussian-smoothed galaxy
density field with a smoothing scale of 4 Mpc. This choice is designed
to probe large-scale structures such as filaments and sheets, which
typically span several megaparsecs in width. We emphasize that our
results pertain to this intermediate scale and may not be directly
comparable to studies that focus on much larger-scale structures
(e.g., superclusters) or smaller-scale environments (e.g., galaxy
groups or cluster cores). Care should be taken when interpreting
similarities or differences across studies that operate at
significantly different spatial resolutions or employ alternative
definitions of cosmic web environments.

The cosmic web serves as the primary source of fuel for galaxy growth,
with its filamentary network playing a crucial role in galaxy
formation and evolution. Our findings emphasize the intricate
interplay between galaxy mass and environment, both of which
collectively shape the evolutionary pathways of galaxies within the
cosmic web.

\section*{ACKNOWLEDGEMENTS}
Authors thank an anonymous reviewer for insightful comments and
suggestions that helped us to improve the draft. BP would like to
acknowledge IUCAA, Pune, for providing support through the
associateship programme. AN acknowledges the financial support from
the Department of Science and Technology (DST), Government of India
through an INSPIRE fellowship. AN thanks Dylan Nelson for help in
downloading and understanding the data from the IllustrisTNG database.

The IllustrisTNG simulations were undertaken with compute time awarded
by the Gauss Centre for Supercomputing (GCS) under GCS Large-Scale
Projects GCS-ILLU and GCS-DWAR on the GCS share of the supercomputer
Hazel Hen at the High Performance Computing Center Stuttgart (HLRS),
as well as on the machines of the Max Planck Computing and Data
Facility (MPCDF) in Garching, Germany.

\section*{Data availability}
The data for IllustrisTNG simulations are publicly available at
https://www.tng-project.org/data/. The data produced in this work will
be shared on reasonable request to the authors.

\appendix 
\section{Appendix}
\label{sec:appen}

We examine how local density and filter width affect our results, with detailed findings for different density bins and filter widths presented in the following two subsections.

\subsection{The impact of varying local density on galaxy fractions}
\label{subsec:locden}

\begin{figure}
    \centering \includegraphics[width =
      1.0\textwidth]{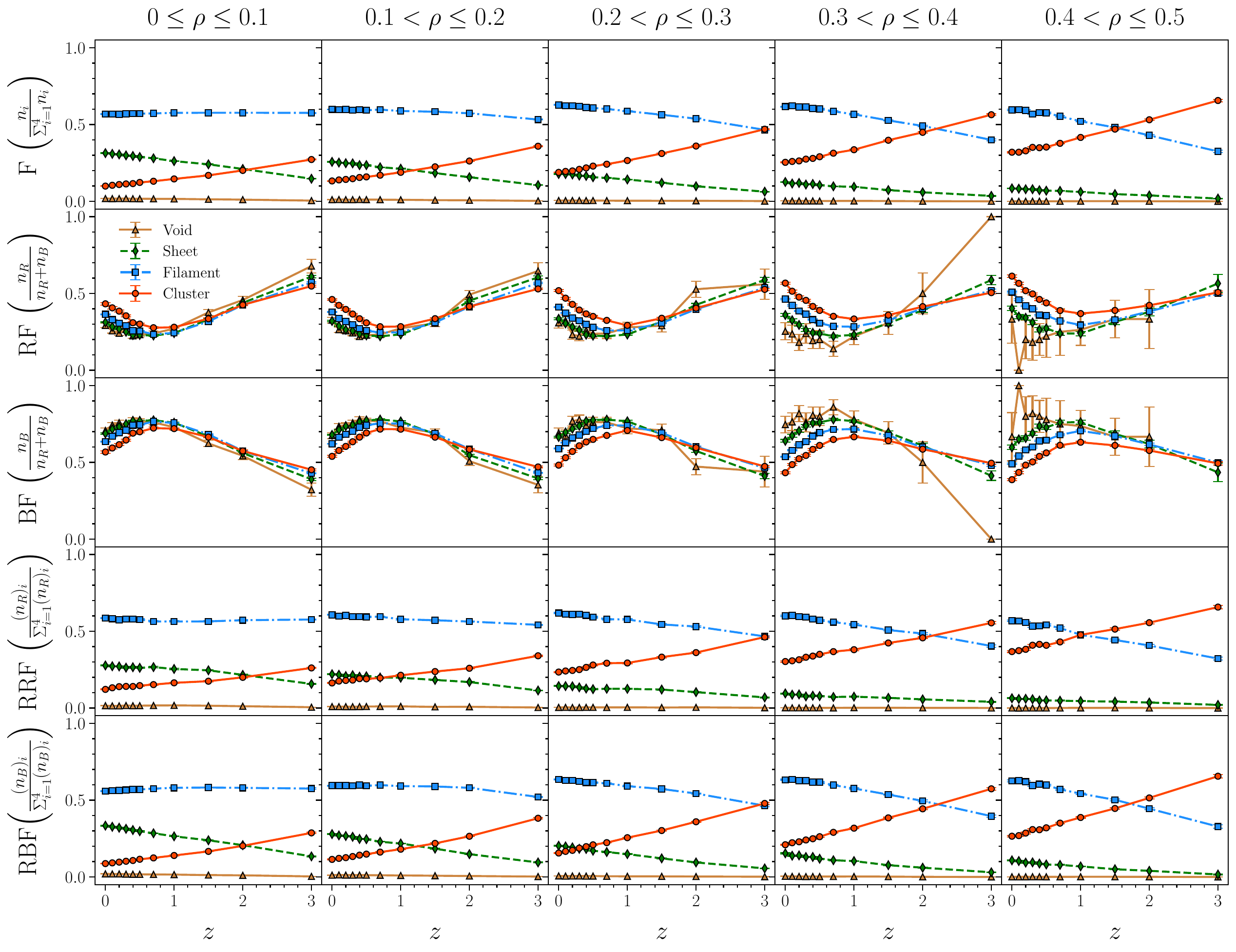}
    \caption{This shows the redshift evolution of
        total fraction (F), red fraction (RF), blue fraction (BF),
        relative red fraction (RRF), and relative blue fraction (RBF)
        across different cosmic web environments. Each row corresponds
        to one specific fraction, while each column shows results for
        a narrow bin in local density, $\rho$, with density ranges
        labeled at the top.}
    \label{fig:denbin}
\end{figure}

We would like to disentangle the relative influence of local density
and cosmic web environment on galaxy colour evolution. We examine the
redshift evolution of total fraction (F), red fraction (RF), blue
fraction (BF), relative red fraction (RRF), and relative blue fraction
(RBF) in different cosmic web environments across 5 narrow local
density bins in \autoref{fig:denbin}. This allows a controlled
comparison of these fractions at fixed densities. Each row presents
the redshift evolution of one fraction, while each column corresponds
to a specific density range. Differences observed across environments
within the same density bin reveal the distinct impact of the
geometric environments of the cosmic web, independent of local
density.

We note that within each local density bin, clear and consistent
differences persist between the cosmic web environments which
indicates that cosmic web geometry influences galaxy evolution
independently of local density. In the top row of
\autoref{fig:denbin}, which shows the total fraction (F) of galaxies
in each cosmic web environment, a consistent pattern emerges across
all density bins. In each local density bin, filaments consistently
host the largest fraction of galaxies, followed by sheets, with voids
and clusters contributing smaller fractions. As we move from low to
high density bins (left to right), the fraction of galaxies in
clusters gradually increases, while the fraction in sheet decreases,
as expected. However, filaments dominate across all bins, regardless
of local density. This suggests that filaments are the most common
environment for galaxies across the full range of densities
examined. The fraction of galaxies in clusters dominates that in
filaments only at higher densities and higher redshifts.

The second row of \autoref{fig:denbin} shows that RF increases over
time across all environments, with the rate of increase depending on
both local density and cosmic web geometry. In all density bins,
clusters consistently exhibit the highest red fractions, followed by
filaments, sheets, and voids, especially at lower redshifts. As local
density increases, overall RF values rise, but the ordering of
environments remains stable, indicating that geometric environment
influences quenching beyond what local density alone accounts for. An
exactly opposite trend is observed in the RF at higher redshifts where
RF is highest in voids followed by sheets, filaments and
clusters. These results imply that both local density and cosmic web
environment shape the build-up of red galaxies, with geometry
modulating quenching efficiency even at fixed density. The third row
of the \autoref{fig:denbin} shows the BF, which is complementary to
the RF and reflects the relative abundance of star-forming galaxies in
each environment and density bin. These results are directly related
to the results presented in the second row of \autoref{fig:denbin} and
support the interpretation that star formation is more sustained in
underdense environments at lower redshift, while it is more efficient
in denser and clustered regions at higher redshifts.

\begin{figure}
    \centering
    \includegraphics[width = 0.8\textwidth]{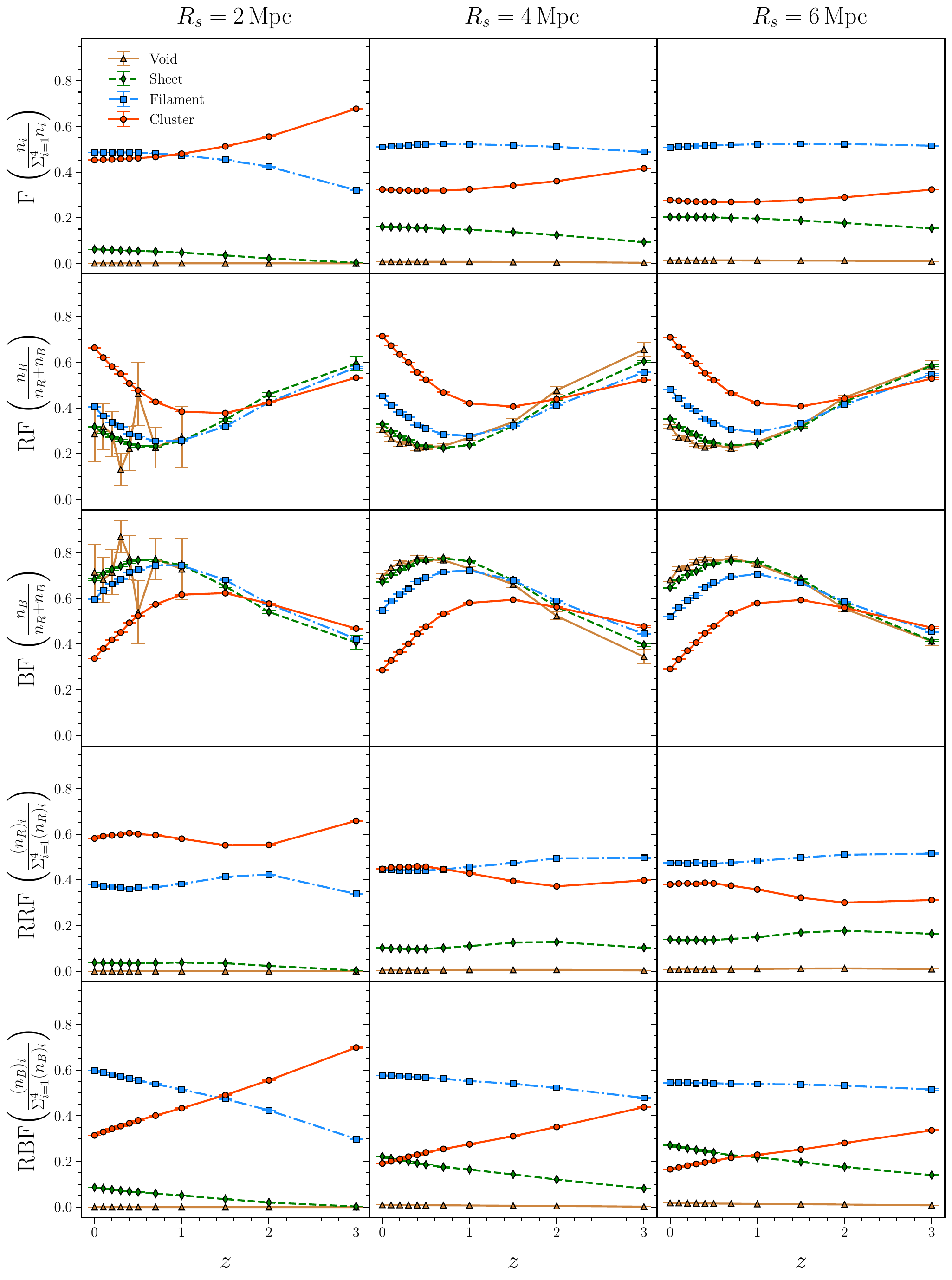}
    \caption{This shows the impact of Gaussian filter width on the
      redshift evolution of total fraction (F), red fraction (RF),
      blue fraction (BF), relative red fraction (RRF), and relative
      blue fraction (RBF) across different cosmic web
      environments. Each row represents one galaxy fraction, while
      each column corresponds to a different filter width.}
    \label{fig:diffsmooth}
\end{figure}

The fourth (RRF) and fifth (RBF) rows of \autoref{fig:denbin} reveal
trends that closely mirror those in the total fraction (F) shown in
the first row, and this similarity provides deeper insight into the
interplay between galaxy population demographics and environmental
influence within fixed local density bins. In all density bins,
filaments consistently dominate the total galaxy population, as shown
in the first row. Correspondingly, in both the RRF and RBF, filaments
also contribute the largest relative share of both red and blue
galaxies, particularly at intermediate densities and redshifts. This
co-dominance arises not necessarily from an intrinsic tendency for
filaments to favour one galaxy type, but from the sheer number of
galaxies they host, positioning them as the primary channel for both
ongoing star formation and quenching within the cosmic web. The
alignment between total fraction (F), RRF and RBF highlights an
important point. Geometric environment shapes not just galaxy
properties, but also where galaxies reside, and these distributions,
in turn, amplify the environment’s role in galaxy evolution. Filaments
are central to this narrative. They act as both the main reservoirs of
galaxies and the transitional sites where galaxies experience a mix of
quenching and star formation. Clusters and voids represent the
extremes, favouring quenching and sustained star formation
respectively. In essence, the combined trends in RRF, RBF, and F
underscore that galaxy colour evolution is not merely about local
density or intrinsic processes, but is profoundly modulated by where
galaxies live in the cosmic web. Filaments, due to their prevalence
and mixed influence, play a pivotal and nuanced role in shaping galaxy
populations. The shifting cross-overs in the first, fourth, and fifth
rows show that the influence of cosmic web environments on galaxy
properties evolves with local density. At low densities, filaments
dominate in both total fraction and relative red and blue fractions,
but as density increases, clusters overtake filaments in hosting red
galaxies, reflecting enhanced environmental quenching. These trends
highlight a dynamic interplay between local density and geometric
environment, where density modulates but does not replace the impact
of cosmic web environment.

\autoref{fig:denbin} clearly demonstrates that galaxy colour evolution
is not solely governed by local density, but is also significantly
influenced by the cosmic web environment. Across all density bins,
distinct differences persist in the red and blue fractions among
voids, sheets, filaments, and clusters. These differences cannot be
explained by density alone. Our results suggest that galaxy colour
evolution is driven by both local density and cosmic web
environment. While local density controls the overall quenching trend,
the geometric environment modulates the efficiency and timing of
quenching, even at fixed local density.

\subsection{The impact of varying filter width on galaxy fractions}
\label{subsec:diffsmooth}
We study the influence of the finite width of the Gaussian filter on
our results in this subsection. The \autoref{fig:diffsmooth} shows
that the filter width used in cosmic web classification significantly
influences the observed galaxy fractions and their evolutionary
trends. While the overall qualitative behaviors persist across
different filter widths, where clusters dominate red fractions, voids
dominate blue fractions, and filaments host the majority of
galaxies. However the relative contributions shift with filter
width. Smaller filter widths better resolve transitional environments
like filaments, crucial for understanding gradual quenching processes.
At smaller widths (2 Mpc), the structures in the cosmic web are more
finely resolved, allowing filaments and sheets to play a prominent
role in hosting both red and blue galaxies.  In contrast, a larger
filter width smooth over finer structures blurring the boundaries
between high-density peaks and surrounding structures. As the filter
width increases, clusters contribute less to the RRF and RBF. This
indicates that clusters become less distinct in the smoothed density
field, likely being diluted into surrounding filaments and sheets due
to loss of high-resolution contrast. Filaments and sheets, in
contrast, gain a larger share of both red and blue galaxies at larger
filter widths These results highlight that filter width is not a
neutral choice. It modulates the apparent dominance of different
environments and thus shapes our interpretation of how galaxies evolve
within the cosmic web.  Therefore, the choice of filter width should
be carefully tailored to the characteristics and resolution of the
dataset being analyzed.

Based on these findings, we adopt a Gaussian smoothing scale of 4 Mpc
for our main analysis. This choice reflects a careful balance between
spatial resolution and structural coherence. A 4 Mpc filter is
slightly below the mean intergalactic separation in our sample ($\sim
5$ Mpc), allowing us to resolve filamentary and sheet-like structures
without introducing excessive fragmentation or noise. At this scale,
transitional environments such as filaments are sufficiently
preserved, enabling a meaningful analysis of gradual quenching and
mixed galaxy populations. In contrast, smaller smoothing scales (e.g.,
2 Mpc) introduce finer detail but can amplify shot noise, while larger
scales tend to over-smooth dense structures, diluting cluster cores
into their surroundings. Moreover, a 4 Mpc filter aligns with the
typical width of filaments observed in both simulations and
large-scale galaxy surveys, ensuring that our results remain
physically interpretable and directly comparable to previous
studies. Overall, this smoothing length provides an optimal trade-off
for capturing the complexity of the cosmic web in a statistically
robust and physically meaningful way.

\end{document}